# Nonlocality and Strength of Interatomic Interactions Inducing the Topological Phonon Phase Transition

Daosheng Tang [1,†]

[1]*School of Physical Science and Technology, Soochow University, Suzhou 215006, China*

**Abstract**

Understanding the phonon behavior in semiconductors from a topological physics perspective provides more opportunities to uncover extraordinary physics related to phonon transport and electron-phonon interactions. While various kinds of topological phonons have been reported in different crystalline solids, their microscopic origin has not been quantitatively uncovered. In this work, four typical analytical interatomic force constant (IFC) models are employed for wurtzite GaN and AlN to help establish the relationships between phonon topology and real-space IFCs. In particular, various nearest neighbor IFCs, *i.e.*, different levels of nonlocality, and IFC strength controlled by characteristic coefficients, can be realized in these models. The results demonstrate that changes in the strength of both the IFCs and nonlocal interactions can induce phonon phase transitions in GaN and AlN, leading to the disappearance of existing Weyl phonons and the appearance of new Weyl phonons. These new Weyl phonons are the result of a band reversal and have a Chern number of ±1. Most of them are located in the $k_z$=0 plane in pairs, while some of them are inside or at the boundary of the irreducible Brillouin zone. Among the various Weyl points observed, certain ones remain identical in both materials, while others exhibit variability depending on the particular case. Compared to the strength of the IFC, nonlocal interactions show much more significant effects in inducing the topological phonon phase transition, especially in cases modeled by the IFC model and SW potential. The larger number of 3NN atoms provides more space for variations in the topological phonon phase of wurtzite AlN than in GaN, resulting in a greater abundance of changes in AlN.

[†]Corresponding author, Email: tangdaosheng1@mail.nwpu.edu.cn

## 1. Introduction

In light of the critical thermal management challenges in electronics and the emergence of nanomaterials, research on phonon heat conduction has experienced significant growth in the past few decades [1-3]. By uncovering the underlying mechanisms of phonon transport in different nanostructures [4-6] and under different extreme conditions, *e.g.*, phonon ballistic-diffusive transport, phonon hydrodynamics, phonon coherences, surface phonon polarization, four-phonon scatterings, phonon-interface transmission, and phonon-electron coupling, various methods have been proposed for specific phonon transport engineering [4, 7-10]. These improvements make it possible to manipulate macro-scale heat conduction through micro-scale "actions", such as tuning thermal transport through the design of nanostructures (superlattices, etc.) and their interface based on the particle and wave nature of phonons. Additionally, it can be realized to regulate phonon transport by inducing various phonon scattering regimes for low-frequency and high-frequency phonons in a targeted manner [8, 9, 11]. However, phonon engineering for enhanced thermal transport is currently still limited, especially in nanostructures and non-equilibrium states, except for decreasing phonon scatterings by increasing the crystal quality and finding materials with high thermal conductivities. Therefore, despite these advances in phonon transport, the exploration of phonon physics continues to deepen towards a better understanding and regulation of heat conduction.

Inspired by the electron-phonon analogy and the experimental observation of the phonon thermal hall effect [12-16], the topological effects of phonons, *i.e.*, the geometrical properties of phonon eigenstates in momentum space, have received much attention [17-20]. In addition to conventional linear polarization, phonons have been observed to exhibit circular polarization, revealing the presence of phonon angular momentum and chirality at the microscopic level [21-26]. Ever since the topological nature of phonons was realized and corresponding theoretical foundations were established, various types of topological phonons have been identified, including topological insulator phonon states, Dirac phonons, Weyl phonons, nodal line phonons, and their complex variations [17, 27-31]. They have been observed in both lattice models and real crystalline materials by first-principles calculations and Neutron diffraction experiments. Similar to the mechanisms in electron systems, the phonon topology is present due to the symmetry-locked or accidental degeneracy (crossing). In specific, topological effects of electrons can result from the presence of Landau level in magnetic fields, inter-hopping between second-nearest neighbors, and spin-orbit coupling [32], those of phonon are expected to be the consequence of specific lattice symmetry as well as their combinations with time reversal symmetry, and the contribution of particular characteristics of interatomic force constants (IFCs). Compared to investigations into topological phonon states in crystalline solids [33-36], a microscopic understanding of topological phonons is still essential [29, 37], especially for three-dimensional materials, which is promising to play a significant role in further studies on thermal transport of topological phonons.

A commonly employed topological phonon model is the phononic Su-Schrieffer-Heeger (SSH) model, featuring a two-atom chain with diverse masses and force constants, and characterized by the winding number or Zak phase [38]. The difference in atomic mass can create a topologically trivial bandgap, whereas the variance in force constants can cause a band inverse, resulting in a nontrivial gap. Upon revisiting this model, one can see that it takes into account only nearest-neighbor interactions, prompting the question of what would occur beyond the nearest-neighbor (1NN) approximation. By introducing third-nearest-neighbor (3NN) interactions, Alisepahi *et al.* [39] found that the winding number fails to describe the topological states in a one-dimensional atom chain, and the Berry connection should be used instead. In metamaterials for airborne sound and phononic crystals, band dispersions can be effectively engineered through nonlocal interactions, *i.e.*, interactions beyond nearest-neighbor atoms [34, 39]. These investigations show the importance of uncovering the microscopic origins of topological phonons in real materials, where nonlocal interactions are prevalent. Actually, more complex nonlocal interactions are present in real crystals, such as three-body interactions and coplanar interactions in addition to two-body pair interactions. In this study, to establish quantitative connections between phonon topology and IFCs, four commonly used analytical IFC models are utilized to model the phonons in three-dimensional lattices, using wurtzite nitrides (GaN and AlN) as illustrative examples. Through topological phonon analyses in the frameworks of various IFC models, this study highlights the impacts of nonlocality and the strength of interatomic interactions in wurtzite nitride semiconductors from an analytical perspective.

## 2. Models and Theories

### 2.1 Interatomic Force Constant Models

The research materials in this work are wurtzite GaN and AlN, both of which are wide bandgap and polar (ionic) crystals. The interactions among atoms in polar crystals can be generally classified into short-range and long-range interactions. And the latter refers to the Coulomb electrostatic forces, which have long-range characteristics. For the short-range term, the detailed interactions differ for different bonding types and configurations, as well as the types of nearest neighbors described as nonlocality here. In this work, we mainly pay attention to the short-range interactions and investigate cases with different levels of nonlocality, including both 1NN and beyond 1NN cases, using typical IFC models and potential functions available in the literature for wurtzite GaN and AlN (Figure 1). The selected IFC models for wurtzite GaN and AlN include Keating's valence force field (KVFF) model [40], the optimized VFF (OVFF) models considering 2NN and coplanar interactions [41-43] and their modified version considering 3NN interactions labelled as KVFF3 and OVFF3 models. Additionally, potential function models including SW [44-46], Tersoff [47-49], SW with a Gaussian term [43], and Tersoff with a Gaussian term are also employed which are all frequently used in molecular dynamics simulations.

In wurtzite structures, for a specific atom, the number of its 1NN, 2NN, and 3NN atoms is generally 4, 12, and 1, respectively [50]. In wurtzite GaN (Figure 1a), this holds true for all scenarios; however, the situation differs in wurtzite AlN (Figure 1b) when considering the 3NN atoms. Specifically, the nearest neighbor conditions vary across different wurtzite structures, depending on their respective lattice constants and internal parameters. In the GaN lattice, there is only one 3NN atom for each cation or anion, which is consistent with the typical wurtzite structure of CdS [50]. However, this type of 3NN atom is not actually the 3NN atom in the AlN structure, as this atom is located among the 2NN atoms. Following the treatment in references [42, 50], this type of atom in AlN is considered as the "3NN" atom in energy calculations, even though it is technically the nominal 3NN atom. Additionally, the real 3NN atoms are also included in the calculations for comparison. With the experimental lattice constant data of wurtzite GaN and AlN, their 1NN, 2NN, and 3NN atoms are listed in detail in the Supplementary Materials.

In the KVFF model, the elastic energy per atom is described by a function of atomic bond lengths and bond angles. For wurtzite lattices, this model can be generalized to

$$\begin{aligned} U_i &= \frac{3\alpha}{16r_0^2}\sum_{j=1}^{3}\left(\mathbf{r}_{ij}^2 - r_0^2\right)^2 + \frac{3\alpha'}{16r_0'^2}\left(\mathbf{r}_{i4}^2 - r_0'^2\right)^2 \\ &+ \frac{3\beta}{8r_0^2}\sum_{j=1}^{3}\sum_{k>j}^{3}\left(\mathbf{r}_{ij}\cdot\mathbf{r}_{ik} - r_0^2\cos\theta_0\right)^2 + \frac{3\beta'}{8r_0r_0'}\sum_{k=1}^{3}\left(\mathbf{r}_{i4}\cdot\mathbf{r}_{ik} - r_0r_0'\cos\theta_0'\right)^2, \end{aligned} \tag{1}$$

where $r_0$ ($r_0'$) is the bond length at the equilibrium state, $\mathbf{r}_{ij}$ is the vector from atom $i$ to $j$, and $\theta_0$ ($\theta_0'$) is the bond angle. The coefficients $r_0'$ and $\theta'$ are introduced to describe the anisotropy of wurtzite lattices. The coefficients $\alpha$ ($\alpha'$) and $\beta$ ($\beta'$) are denoted as the bond-stretching and bond-bending constants. The coefficients used for this model are from reference [40], where data fitting was performed based on the experimental elastic properties. In general, the KVFF model only considers 1NN interactions (the first two terms) and three-body interactions using a bending constant coefficient (the last two terms) to some extent.

The OVFF model, a modification of the VFF model [41], incorporates factors such as the 2NN interaction, bending-stretching interaction, coplanar (COP) interaction terms, and lattice anisotropy. In this model, the elastic energy for each atom is calculated as

$$\begin{aligned} U_i &= \frac{3}{16r_0^2}\sum_{j}^{\mathrm{FN}(i)}\alpha_{ij}\left(\Delta r_{ij}\right)^2 + \frac{3}{16}\sum_{m}^{\mathrm{SN}(i)}\mu_{ij}\left(\Delta r_{im}\right)^2 \\ &+ \frac{3}{8}\sum_{j}^{\mathrm{FN}(i)}\sum_{k>j}^{\mathrm{FN}(i)}\left\{\beta_{jik}\left(\Delta\theta_{jik}\right)^2 + \gamma_{jik}\Delta r_{ij}\Delta\theta_{jik} + \xi_{jik}\Delta r_{ij}\Delta r_{ik} + \sum_{l}^{\mathrm{COP}(jikl)} v_{ijkl}\Delta\theta_{jik}\Delta\theta_{ikl}\right\}, \end{aligned} \tag{2}$$

where

$$\begin{aligned} \Delta r_{ij} &= \left(r_{ij}^2 - d_{ij}^2\right)/d_{ij}, \\ \Delta\theta_{jik} &= \left(\mathbf{r}_{ij}\cdot\mathbf{r}_{ik} - d_{ij}d_{ik}\cos\theta_{jik}\right)/\sqrt{d_{ij}d_{ik}}. \end{aligned} \tag{3}$$

The abbreviations FN($i$) and SN($i$) represent all 1NN and 2NN atoms, respectively. The coefficients $\alpha$, $\mu$, $\beta$, $\gamma$, $\xi$, and $\upsilon$, are used to describe the 1NN bond stretch (the first term), 2NN bond stretching (the second term), bond bending (the third term), stretch-bend interaction (the fourth term), cross-stretch interaction (the fifth term) and coplanar interaction (the last term). Specifically, to describe the anisotropy, the first-nearest neighbor bond lengths are labelled as $d_1$ and $d_{1c}$, and the second-nearest neighbor bond lengths are $d_2$ and $d_{2c}$, where "$c$" denotes the interaction parallel to the $c$ direction. The bond angles are represented as $\theta$ and $\theta_c$. The first and third terms in the OVFF model are almost the same as those in the KVFF model, and the optimized parts mainly refer to the inclusion of 2NN and coupling interaction terms. The coefficients in this model are determined by the fitting procedure based on the experimental phonon dispersion data [41]. Beyond the 1NN and 2NN interactions, the 3NN interaction is modeled by inserting a two-body 3NN term [42, 50] with IFC coefficient $\delta$ into the KVFF and OVFF models.

Despite their effectiveness in forecasting elastic and phonon behaviors in solids, VFF models necessitate more sophisticated potential functions for intricate processes with significant atomic displacements, such as phase transitions [51]. These potential functions provide more physical models to investigate the topological phonon states in wurtzite crystals. For wurtzite semiconductors GaN and AlN, there are widely used potential functions such as the Stillinger-Weber (SW) and Tersoff potentials. These potentials successfully model anisotropic directional bonding by including three-body interactions. The SW model includes both the two-body and three-body terms explicitly.

$$U = \sum_i \sum_{j>i} \phi_2\left(r_{ij}\right) + \sum_i \sum_{j\neq i} \sum_{k<j} \phi_3\left(r_{ij}, r_{ik}, \theta_{ijk}\right), \tag{4}$$

where

$$\begin{aligned} \phi_2\left(r_{ij}\right) &= A_{ij}\varepsilon_{ij}\left[B_{ij}\left(\frac{\sigma_{ij}}{r_{ij}}\right)^{p_{ij}} - \left(\frac{\sigma_{ij}}{r_{ij}}\right)^{q_{ij}}\right]\exp\left(\frac{\sigma_{ij}}{r_{ij}-a_{ij}\sigma_{ij}}\right), r/\sigma < a \\ \phi_3\left(r_{ij}, r_{ik}, \theta_{ijk}\right) &= \lambda_{ijk}\varepsilon_{ijk}\left[\cos\theta_{ijk} - \cos\theta_{0ijk}\right]^2 \exp\left(\frac{\gamma_{ij}\sigma_{ij}}{r_{ij}-a_{ij}\sigma_{ij}}\right)\left(\frac{\gamma_{ik}\sigma_{ik}}{r_{ik}-a_{ik}\sigma_{ik}}\right), r/\sigma < a \end{aligned} \tag{5}$$

The cutoff $a$ is used to determine the interaction range. The coefficients in this model include $\varepsilon$, $\sigma$, $A$, $B$, $p$, $q$, $\lambda$, $\gamma$, and $\cos\theta_0$. In the standard SW potential, only the 1NN interactions are involved, while the 3NN terms, responsible for the $c/a$ ratio, are not included [43]. Following Bera's method, a Gaussian term

$$U = \frac{H}{\sigma_h\sqrt{2\pi}}\exp\left[-\left(\frac{r_{ij}-r_{mh}}{\sqrt{2}\sigma_h}\right)^2\right] \tag{6}$$

can be added to describe the 3NN interaction as a modification to the SW potential. In this formula, the coefficient $H$ determines, together with the standard deviation $\sigma_h$, the peak height of the Gaussian function, and $r_{mh}$, the peak position. More clearly, $r_{mh}$ is the center of the Gaussian function, which is situated between the second and third nearest neighbor distances, and $\sqrt{2}\sigma_h$ is the width of the Gaussian function. In the original form of this Gaussian term in reference [43], the coefficients of this exponential function include $\varepsilon$ and $C$. In the following discussion, the coefficient $\varepsilon$ is varied to tune the strength of this 3NN interaction.

Another important 3-body potential function is the Tersoff potential. In this potential, the total energy of the lattice system is written in the form of two-body potentials,

$$
\begin{aligned}
U &= \frac{1}{2}\sum_i \sum_{j\neq i} V_{ij} \\
V_{ij} &= f_C\left(r_{ij}+\delta\right)\left[ f_R\left(r_{ij}+\delta\right) + b_{ij} f_A\left(r_{ij}+\delta\right)\right].
\end{aligned} \tag{7}
$$

where

$$
\begin{aligned}
f_C(r) &= \begin{cases} 1 & r < R-D \\ \frac{1}{2}-\frac{1}{2}\sin\left(\frac{\pi}{2}\frac{r-R}{D}\right), & R-D < r < R+D \\ 0 & r > R+D \end{cases} \\
f_R(r) &= A\exp\left(-\lambda_1 r\right), \\
f_A(r) &= -B\exp\left(-\lambda_2 r\right), \\
b_{ij} &= \left(1+\beta^n \zeta_{ij}^n\right)^{-\frac{1}{2n}}, \\
\zeta_{ij} &= \sum_{k\neq i,j} f_C\left(r_{ik}+\delta\right) g\left[\theta_{ijk}\left(r_{ij}, r_{ik}\right)\right] \exp\left[\lambda_3^m \left(r_{ij}-r_{ik}\right)^m\right], \\
g(\theta) &= \gamma_{ijk}\left(1+\frac{c^2}{d^2}-\frac{c^2}{\left[d^2+\left(\cos\theta-\cos\theta_0\right)^2\right]}\right).
\end{aligned} \tag{8}
$$

In Eq. (8), the coefficients A and B denotes the strength of the repulsive and attractive interactions. The 3-body interaction is hidden in the attractive interaction term with a coefficient $\zeta_{ij}^n$ that mainly depends on the angles among three atoms. For simplicity, detailed explanations of the coefficients in these models are not present. More details including the coefficients and lattice constants used in these models can be found in the corresponding references [44, 45, 47, 49, 52] and the Supplementary Materials. Actually, the force constant models and potential functions for wurtzite lattice structures are far more than those [53-58]. In this work, the selection of IFC models mainly concerns their availability for better comparisons to illustrate the topic of this work, *i.e.*, the effects of nonlocality and strength on the topological phonon phase transition.

### 2.2 Topological Phonon Theory

The primary objective of this research is to explore the Weyl phonon properties of wurtzite nitrides. By means of IFC models, phonon dispersions and eigenstates can be determined through the application of lattice dynamic theory. The characteristic topological physical quantity, known as the Chern number, is defined as the integration of the phonon Berry curvature over a closed surface. It is calculated using the Wilson-loop method, based on the phonon eigenstate data around the crossing points. In the calculations for phonon topology, tight-binding Hamiltonian of phonons are adopted which is converted from the force constants,

$$H = \frac{1}{2}\sum_{l,m,\alpha}\sum_{l',m',\beta}\Phi_{\alpha\beta}\left(R_{ml} - R_{m'l'}\right)u_{lm}^{\alpha}u_{l'm'}^{\beta} \tag{9}$$

In this formula, the force constant matrix is denoted by $\Phi$, and the parameters $l$, $m$, and $\alpha/\beta$ are used to label the number of unit cells, atoms in a unit cell, and directions, respectively. Through the examination of all the crossing phonon points in the irreducible Brillouin zones (IBZs), the Weyl phonon property of each nitride can be determined. To display the nontrivial topological characteristics, the Berry curvatures at the corresponding $k$-planes, as well as the phonon surface states projected onto the surface BZs, are also calculated. More details on topological phonon theory and calculations can be found in references [30, 36, 38].

## 3. Results and Discussion

### 3.1 Weyl phonons in original models

The phonon dispersions of GaN and AlN were calculated using the IFC models and potential functions, as depicted in Figure 2 and Figure 3. Despite variations in specific magnitudes, all four models agree on the presence of a significant band gap in GaN between high- and low-frequency phonons [20, 59]. Broadly speaking, the acoustic branches of the four GaN models display consistent phonon behavior, with variations primarily evident in the optical branches, particularly in the high-frequency optical branches. The dispersions of AlN illustrated in Figure 3 show consistent predictions between the Keating and Tersoff models, in contrast to the significant deviations observed with the SW model.

Investigation into the Weyl phonon properties of GaN and AlN is carried out by employing diverse IFC models. In Table 1, the calculation results are presented, showing the assignment of a trivial or nontrivial label for each case and the corresponding number (in parentheses) of Weyl phonons located within the irreducible Brillouin zone (IBZ). It can be found that GaN exhibits nontrivial properties in the KVFF, OVFF, and SW models, with four, two, and two Weyl phonons in the IBZ, respectively. In detail, there are two pairs of Weyl phonons related to bands 1 and 4 in the KVFF model. In the OVFF model, there is only one pair of Weyl phonons with band 4 and one pair of Weyl phonons with band 1 in the SW model. Slightly different from that in GaN, the AlN modelled by the SW model exhibits nontrivial Weyl phonons with two pairs in the IBZ. For both GaN and AlN systems, the Tersoff model yields trivial Weyl phonon properties.

Figure 4 further depicts the detailed phonon band crossing at the Weyl points, accompanied by corresponding topological features including Wannier charge center (WCC) evolutions, Berry curvature, and arc surface states in Figure 5-Figure 7. It can be found that the nontrivial Weyl points result from the crossing of either bands 1 and 2 (sorted by frequency) in the acoustic branches or bands 4 and 5 in the optical branches. The Weyl phonon in GaN, predicted by the OVFF model, is slightly different from the other three (type-I Weyl phonons) as it is a type-II Weyl phonon. The phonon band connections are further illustrated with different colors based on the phonon eigenvectors [60]. And for the Weyl phonons at band 1, detailed results show that these types of crossing points are formed due to the crossing of three bands. It is also worth noting that GaN and AlN do not show a significant difference in the simple KVFF models, and their nontrivial phonon crossing points in the SW model are also very similar. While several Weyl phonons host obvious arc surface states projected along the [0001] direction, the other surface states are not clean enough due to the overlap with bulk states as shown in Figure 7. Combining the results in Table 1 and Figure 4, it is evident that a phonon phase transition is promising to occur in wurtzite nitrides under this framework. Specifically, by tuning the strength of the IFC and introducing nonlocal interatomic interactions, current Weyl phonons may disappear and new Weyl phonons may be generated.

### 3.2 Weyl phonons in modified models

While Weyl phonons, as well as their phase transitions, have been reported in various wurtzite materials, the origins of Weyl phonons are still not clear [27-29, 36]. In this section, based on the frameworks of IFC models and potential functions, Weyl phonons are searched for in modified models with varying atomic interaction coefficients and nonlocality beyond 1NN interactions. The atomic interactions discussed here are classified as the strength and nonlocality of IFCs. These include bond-stretching and bond-bending interactions, bending-stretching coupling interactions, coplanar interactions, and 2NN and 3NN interactions, as shown in Table 2.

- **Strength**

For the KVFF model, the coefficients accounting for the IFC strength are the 1NN bond-stretching coefficient $\alpha$ and the bond-bending coefficient $\beta$. Consequently, these coefficients are modified by a reasonable variation (*e.g.*, ±5% for the 1NN stretching) to mimic varying degrees of IFC strength. Within the variation in 1NN bond-stretching for GaN, the optical phonon branches, especially those with high frequencies, change significantly, while the acoustic ones remain nearly constant. When the strength of bond-bending changes, both the optical and acoustic phonon branches are affected, although the magnitude of the changes is smaller compared to those observed in cases with varying bond-stretching. In AlN, the general changes are nearly the same, but the corresponding change magnitude is noticeably larger. For the OVFF model, coefficients for 2NN bond-stretching $\mu$ and stretching-bending interactions ($\beta\gamma\xi$) are present in addition to those in the KVFF model. To make the changes in phonon dispersion more significant, larger degrees of variation are adopted for these coefficients, with the exception of inducing possible phonon phase transitions. It can be seen from the phonon dispersions that both the acoustic and optical branches vary significantly with changes in the 2NN bond-stretching and bending-stretching coupling.

The coefficients in potential functions are not as simple and clear as those in VFF models. In the SW model, two coefficients are selected to introduce variations, including the coefficient $A$ in the two-body term and $\lambda$ in the three-body term. The results in phonon dispersion further confirm that variations in two-body 1NN IFC strength mainly induce changes in the optical branches, while variations in three-body IFC strength induce changes in both the acoustic and optical branches. In the Tersoff model, there are no coefficients for two-body or three-body terms. Instead, the coefficients accounting for repulsive and attractive interactions are selected. Under the conditions of varying coefficients for repulsive ($A$) and attractive ($B$) interactions, both the acoustic and optical branches change significantly. Also, the lattice parameters change during the variations of coefficients, which are reflected in the changes of the Brillouin zone. There is also a slight difference in the case of the Tersoff model. The phonon frequencies of AlN decrease with both an increase and a decrease in the attractive coefficient $B$. It should be noted that the lattice parameters are kept constant in the IFC-model-based calculations, while they are optimized in the potential-function-based calculations. This is because there is a minimum energy within the framework of potential functions. More details on phonon dispersions with varying IFC strength can be found in the Supplementary Materials. As a matter of fact, besides the cases discussed in this work, there are far more variations in coefficients than those mentioned. These variations also include coefficients that consider anisotropy and different atomic pairs. Also, the coefficients in the OVFF model and potential functions mentioned above can be explicitly changed for different directions and atomic pairs.

By tuning the coefficients for IFC strength, tens of different cases are generated for phonons in GaN and AlN, where the phonon topology also exhibits different characteristics, as shown in Table 3. Nontrivial variations, *i.e.*, phonon phase transitions, are present in the enlarged phonon dispersions as shown in Figure 8 and Figure 9, which display the Weyl phonons induced by variations in the IFC strength coefficients of the IFC models and potential functions, respectively. For a clearer understanding, the positions of Weyl points are all illustrated in the IBZ in Figure 10. In the KVFF model, no phonon phase transition is found in either GaN or AlN. Two pairs of Weyl phonons at bands 1 and 4 in the IBZ remain present in both the original models and the modified models with different strengths of IFCs. In the OVFF model, additional Weyl phonons are generated in several cases, while the original Weyl points disappear in the other cases, except for the case with increasing bond-bending and bending-stretching interactions (($\beta$, $\gamma$, $\xi$)=1.2($\beta$, $\gamma$, $\xi$)$_0$), where no phase transition occurs. It can be seen from Table 3 and Figure 8 that the Weyl phonons in the original OVFF model arise from the crossing of bands 4 and 5. Newly generated Weyl phonons result from the crossing of bands (8,9), (8,9,10,11), (8,9,10), and (7,8,9) in the case of $\alpha$=1.05$\alpha_0$, $\alpha$=0.95$\alpha_0$, $\mu$=1.20$\mu_0$, and ($\beta, \gamma, \xi$)=0.8($\beta, \gamma, \xi$)$_0$, respectively. While their basic profiles are similar, the relative relationships between the bands differ. The newly generated Weyl phonons from the crossing of bands 8 and 9 are present in the case with $\alpha$=1.05$\alpha_0$. In cases where $\alpha$=0.95$\alpha_0$, $\mu$=1.20$\mu_0$, and ($\beta\gamma\xi$)=0.80($\beta\gamma\xi$)$_0$, newly generated Weyl phonons are also located around the frequency of 900 cm$^{-1}$. However, their detailed formations are different. In the first case, the Weyl point consists of the crossing of four bands, including bands 8-11. In the second case, it results from the crossing of three bands, including bands 8-10. For the last case, band crossings occur among bands 7-9. In real space, the nontrivial phonon topology normally corresponds to rotational atomic vibrations or circular phonon polarizations. More detailed information can be found in the Supplemental Materials, where the atomic vibrations of Weyl phonons are visualized.

In the framework of SW and Tersoff potentials, new Weyl phonons are generated in the cases of $A$=0.8$A_0$ and $\lambda$=1.2$\lambda_0$ in both GaN and AlN. For GaN, in the cases of $A$=0.8$A_0$ and $\lambda$=1.2$\lambda_0$, *i.e.*, the decrease in the two-body interaction and increase in the three-body interaction, Weyl points are generated due to the variations in the strength of the IFC, resulting in the crossing of bands 4 and 5. There are no additional Weyl points in the other cases. In the scenario where the coefficient $A$ equals 0.8$A_0$ in AlN, new Weyl points emerge due to the band crossing of bands 8 and 9. Moreover, when $\lambda$ equals 1.2$\lambda_0$, two pairs of Weyl points are triggered at bands 9 and 11. Different from that in GaN, the increase in the two-body interaction ($A$=1.2$A_0$) and the decrease in the three-body interaction ($\lambda$=0.8$\lambda_0$) break the nontrivial properties of band 1. As shown in the enlarged phonon dispersions (Figure 9) and positions in the IBZ (Figure 10), it can be found that the Weyl phonons formed by the same bands in GaN modeled by the SW potential are similar. For example, the Weyl points at band 1 formed by the crossing among bands 1-3 are present in both cases $A$=0.8$A_0$ and $\lambda$=1.2$\lambda_0$. And with the variations of IFC strength, the Weyl points at band 4 remains present with nontrivial property in four different cases. Variations of IFC strength induce more Weyl phonons in AlN than those in Ga, such as Weyl points at bands 9 and 11, which are not present in the cases modeled by the VFF models.

Besides, Weyl points from the crossing of bands 11 and 12 are located at the top boundary of the IBZ, while all other Weyl points in the SW model are all present at $k_z$=0 plane (Figure 10). Under the conditions of the Tersoff potential, AlN is always a trivial phonon system, both in the original model and in the model with varying coefficients, regardless of how the coefficients change. Weyl phonons are generated in cases of varying $A$ and increased $B$, while the original GaN system is topologically phonon-trivial. The newly generated Weyl phonons are not all the same. Those in the case of increased $A$ are located at the $k_z$=0 plane in the IBZ while the others are not. The Weyl points resulting from the intersection of bands 9 and 10, as well as the intersection of bands 10 and 11, are classified as type-II Weyl phonons and are not shown in cases with other IFC models and potentials. By varying the IFC strengths in different IFC models, where the IFCs are limited within the 1NN distance, the positions and presence of Weyl phonons are altered correspondingly. Specifically, Weyl phonons can be present in both the acoustic and optical branches, which are not only located in the $k_z$=0 plane, but also inside the IBZ and on the boundary of the IBZ. Additionally, Weyl phonons are generally different among different models as displayed in the enlarged phonon dispersions, indicating the significance of the local atomic bonding configurations.

- **Nonlocality**

In the original VFF and potential models, only 1NN interactions are generally considered [45, 49, 61]. The 2NN and interactions among neighboring farmer atoms are much smaller in magnitude and are often neglected in phonon and elastic calculations. Indeed, nonlocal terms beyond 1NN interactions will not significantly alter the phonon dispersion profiles and magnitudes (see Supplementary Materials). However, the topological effects of phonons may be sensitive to these nonlocal interatomic interactions, which are investigated in this section.

In the KVFF and OVFF models, a 3NN interaction term is incorporated, featuring diverse interaction strengths, denoted as the VFF3 models. In contrast to the fluctuations observed in instances characterized by varying strength coefficients, the variations in phonon dispersions show minimal overall increase or decrease. Instead, slight changes mainly occur in the dispersion profiles. For example, in cases with large 3NN interactions, the phonon dispersion at large $k_z$ vectors shows the most significant changes. It should be noted that the positions of 2NN and 3NN atoms in wurtzite GaN and AlN are not exactly the same, as mentioned in section 2. For both structures, there are 12 2NN atoms around each atom with the same element, where 6 of them lie in the same horizontal plane (the plane perpendicular to the *c*-axis) and the other 6 lie above or below it. For GaN, there is only one 3NN atom for each atom with a different element, and it lies directly above or below the current atom with determined positions. With significantly different lattice constants and internal parameters, the corresponding 3NN atom in AlN is the nominal 3NN atom, which is actually located among 12 2NN atoms. There are 9 real 3NN atoms according to the neighbor distance, which are the real 3NN atoms for the AlN structure. In the OVFF model, the 2NN interaction term has been included in its original form, as well as the coplanar interaction. To examine the effects of nonlocal terms, we also explore cases without the 2NN and COP terms, as well as cases with varying strengths of the 3NN term. Under the framework of the OVFF model, phonon dispersions change significantly after subtracting the 2NN or COP term. For cases using the SW and Tersoff potentials, a Gaussian term is included to model the 3NN interaction, with the center of the Gaussian function located between the 2NN and 3NN atoms. Different from the calculations of VFF models, the 3NN atoms are determined by searching for the actual distance directly, which are the real 3NN atoms. The introduction of 3NN interactions results in changes to both the acoustic and optical branches. For cases with 3NN interactions, different strengths of 3NN interactions are also adopted to capture the detailed variations in phonon phase. Table 4 lists all the cases in this work for investigating the effects of nonlocal interaction terms on the topological phonon phase.

In the KVFF model, the phonon phase transition in GaN is not initiated by 3NN interactions until the strength of the 3NN interaction reaches five times its original value. In detail, the Weyl points at band 1 disappear when the 3NN interaction coefficient $\delta$ is equal to $5\delta_0$. In contrast, both the nominal and real 3NN interactions induce additional Weyl phonons in AlN. With both nominal and real 3NN interactions, the Weyl points at band 1 are absent in AlN. However, the nominal 3NN interaction of strength $\delta$=$5\delta_0$ and the real 3NN interaction of strength $\delta'$=$2\delta'_0$ both trigger new Weyl points at band 9, and bands 9 and 10, respectively. In the OVFF model, the Weyl phonon properties remain constant despite varying 3NN interactions, even though the strength has increased to twenty times its original value. In the framework of SW and Tersoff potentials, 3NN interaction terms can definitely induce the phonon phase transition. For the GaN structure modeled by the SW potential, both 3NN interactions can result in phonon phase transitions. The smaller 3NN interactions ($\varepsilon$=0.1$\varepsilon_0$ and $\varepsilon$=$\varepsilon_0$) lead to the loss of Weyl points at band 1, while the larger 3NN interactions ($\varepsilon$=2$\varepsilon_0$ and $\varepsilon$=10$\varepsilon_0$) induce new Weyl phonons at band 4. For the AlN structure, new Weyl points are present at bands 9 and 10 due to the large 3NN interactions. The GaN and AlN structures modeled by the original Tersoff potentials are trivial for phonon topology. Both of them become nontrivial with 3NN interactions. For GaN, a small 3NN interaction induces two pairs of Weyl points in bands 9 and 10, while a large one induces one pair at band 9 only. For AlN, both the standard and medium 3NN interactions induce one pair of Weyl points at band 4.

Phonon dispersions in original models for GaN and AlN and in modified models with 3NN interactions are displayed in Figure 11 and Figure 12. As illustrated by the enlarged Weyl points in Figure 11, the Weyl phonons resulting from the crossing of bands 4 and 5 are present in original and modified models. And this type of Weyl phonons is similar in GaN and AlN structures. The Weyl points in GaN modeled by the KVFF model at band 1 are actually the crossing among bands 1-3, located at $k_z$=0 plane (Figure 13), which disappear with large 3NN interaction. And this type of Weyl phonons is much different in GaN and AlN structures. For AlN modeled by the KVFF model, both nominal and real 3NN interactions with large magnitudes induces the Weyl points at band 10 which are very similar and result from the crossing of bands 10-12, as shown in Figure 11(f) and (g). However, their positions in the IBZ are significantly different. The Weyl points in the case of large real 3NN are present inside the IBZ with large $k_z$ components while those in the case of large nominal 3NN are located at the $k_z$=0 plane (Figure 13). The Weyl points at band 9 consisting of the crossing of bands 9 and 10 are only shown in the case of large real 3NN interaction (Figure 11(i)).

When modeled with the SW potential incorporating a Gaussian-type 3NN term with both large and medium strengths ($\varepsilon$=2$\varepsilon_0$ and $\varepsilon$=10$\varepsilon_0$), the Weyl phonons in GaN demonstrate similarities to the phonon characteristics formed by the crossings of bands (1,2) and (4,5) in AlN, as shown in Figure 12(a)-(c). With the 3NN interactions of large and medium levels, Weyl points at band 4 are induced in GaN. However, the Weyl points in original model disappear with smaller 3NN interactions. In the original AlN system, two pairs of Weyl phonons emerge at the crossing point, located at the $k_z$=0 in the IBZ. One of them is formed by the bands 1 and 2, and the other by the bands 4 and 5. These Weyl points remain present in cases with 3NN interactions of small strength while the former one is lost when the strength of 3NN interaction increases to two times its initial value. With the continue increase of the 3NN interaction strength, Weyl points at bands 9 and 10 are generated and the Weyl points at band 4 also changes significantly in profiles as depicted in Figure 12. In the framework of Tersoff potential, both GaN and AlN are topologically nontrivial for phonons with their original coefficients. Interestingly, Weyl phonon phase transitions occur in both two materials but with different strength levels. In GaN simulated with the Tersoff potential, Weyl points at bands 9 and 10 are present in the case of small 3NN interaction ($\varepsilon$=0.1$\varepsilon_0$), and only the Weyl points at band 9 is present in the case of large interaction ($\varepsilon$=10$\varepsilon_0$). Also, these three pairs of Weyl points are not the same, even for the two pairs at band 9 (Figure 12). They vary in dispersion profiles and positions within the IBZ, as depicted in Figure 11 and Figure 12. One pair of Weyl points, generated by the small 3NN interaction, is present in the $k_z \neq 0$ plane, while the other Weyl points are located at the $k_z = 0$ plane.

**3.3 Characteristics during Weyl phonon transitions**

To uncover the cause of the topological phonon phase transitions observed with changes in IFCs, we employ band connection results to demonstrate the conditions for band inversion. The results illustrate that nearly all Weyl phonon transitions are accompanied by band inversions at the Weyl points, including changes in the order of bands and changes in the types of band crossings (for crossings of three or more bands). Selected cases with both phonon dispersions and corresponding Berry curvature profiles are depicted in Figure 14. When a small real 3NN interaction is added, the corresponding point changes to the phonon crossing of bands 1 and 2. When a small real 3NN interaction is added, the corresponding point changes to the phonon crossing of bands 1 and 2(Figure 14(b)). The corresponding Berry curvature at band 1 becomes trivial without any source or sink features (Figure 14(d)). For the other two pairs of compared cases shown in Figure 14(e)-(l), the band orders are reversed during the phonon phase transitions. More details for other cases can be found in the Supplemental Materials.

It is worth noting that in the search process, some Weyl points could be inadvertently omitted if they are positioned very closely together or if their nontrivial properties are not readily apparent.

## 4. Conclusions

In this work, four typical IFC models including VFF models and three-body potential functions are used to model the phonon dispersions and topological effects in wurtzite GaN and AlN structures concerning the topological phonon phase transition by the nonlocality and strength of IFCs. The main conclusions are as follows.

(1) Both GaN and AlN are topologically phonon nontrivial in the IFC models and SW potential, while they are both trivial system when modeled by the Tersoff potential. In the simplified model where only nearest neighbor interactions are considered, Weyl phonons in GaN and AlN exhibit nearly identical characteristics. Changes in the strength of the IFCs and nonlocal interactions can induce a phonon phase transition in GaN and AlN, leading to the disappearance of existing Weyl phonons and the appearance of new Weyl phonons.

(2) New Weyl phonons from the variations of IFCs are found to have a Chern number of ±1, supported by the results of Wannier charge center evolution and Berry curvature. Most of them are located in the $k_z$=0 plane in pairs, while some of them are inside and at the boundary of the Brillouin zone or appear singly. Among the various Weyl points observed, certain ones remain identical in both materials, showing the common characteristic of GaN and AlN as wurtzite structures, while others exhibit variability depending on the particular case.

(3) Compared to the strength of the IFC, nonlocal interactions show much more significant effects in inducing the topological phonon phase transition, especially in cases modeled by the IFC model and SW potential. When considering the nonlocal interatomic interactions, GaN and AlN differ significantly in their 3NN atoms, in addition to their atomic masses and lattice constants. The larger number of 3NN atoms provides more space for variations in the topological phonon phase of wurtzite AlN than in GaN, resulting in a greater abundance of changes in AlN. Examination of bands, through their connections, reveals that alterations in IFCs lead to band reversals, consequently triggering topological phonon phase transitions.

This study confirms that a feasible three-dimensional topological phonon model can be created for wurtzite nitride structures using various analytical IFC models. This provides more opportunities for further investigation to uncover the microscopic origin of topological phonons.

## Supplementary Materials

See more details about IFC models, potential functions, nearest-neighbor atoms, and phonon dispersions in cases with different coefficients in the Supplementary Materials. Additionally, the Supplementary Data contains information on the Berry curvature, Wannier charge center evolution, and eigenstates.

## Acknowledgements

This work was financially supported by the National Natural Science Foundation of China (No. 52206105), Jiangsu Funding Program for Excellent Postdoctoral Talent (No. 2022ZB594), the China Postdoctoral Science Foundation (No. 2021M702384), and Natural Science Foundation of the Jiangsu Higher Education Institutions of China (No. 22KJB470008).

## Conflict of Interest

The authors have no conflicts to disclose.

## Data availability

The data supporting the findings of this study are available from the corresponding author upon reasonable request.

Table 1. Properties of GaN and AlN regarding Weyl phonons. The numbers in brackets denote the number of Weyl phonons in the IBZ and their corresponding band index.

| Material \ Model | KVFF | OVFF | SW | Tersoff |
|---|---|---|---|---|
| GaN | *Nontrivial* (4, 1,4) | *Nontrivial* (2, 4) | *Nontrivial* (2, 1) | Trivial |
| AlN | *Nontrivial* (4, 1,4) | - | *Nontrivial* (4, 1,4) | Trivial |

Table 2. Variable coefficients and terms

| Types | Coefficients | Models (comparisons) |
|---|---|---|
| Bond-stretching | 1NN bond stretching | KVFF |
| | 2NN bond stretching | OVFF |
| Three-body interaction | Bond-bending | VFFs |
| | Bending-stretching coupling | VFFs |
| Nonlocal interaction | 2NN bond stretching | OVFF vs. OVFF without 2NN |
| | Coplanar interactions | OVFF vs. OVFF without COP |
| | 3NN bond stretching | KVFF vs. KVFF3 |
| | | OVFF vs. OVFF3 |
| | | SW vs. SW + Gaussian |
| | | Tersoff vs. Tersoff + Gaussian |

Table 3. Weyl phonon properties of GaN and AlN with varying coefficients in strength. The numbers in parentheses denote the number of Weyl phonons in the IBZ and their band index.

| Models \ Materials | Variations of coefficients | GaN | AlN |
|---|---|---|---|
| KVFF | Original | *Nontrivial* (4, 1,4) | *Nontrivial* (4, 1,4) |
| | Stretching $\alpha$=1.05$\alpha_0$ ↑ | *Nontrivial* (4, 1,4) | *Nontrivial* (4, 1,4) |
| | Stretching $\alpha$=0.95$\alpha_0$ ↓ | *Nontrivial* (4, 1,4) | *Nontrivial* (4, 1,4) |
| | Bending $\beta$=1.05$\beta_0$ ↑ | *Nontrivial* (4, 1,4) | *Nontrivial* (4, 1,4) |
| | Bending $\beta$=0.95$\beta_0$ ↓ | *Nontrivial* (4, 1,4) | *Nontrivial* (4, 1,4) |
| OVFF | Original | *Nontrivial* (2, 4) | - |
| | 1NN Stretching $\alpha$=1.05$\alpha_0$ ↑ | *Nontrivial* (4, 4,8) | - |
| | 1NN Stretching $\alpha$=0.95$\alpha_0$ ↓ | *Nontrivial* (4, 4,9) | - |
| | 2NN Stretching $\mu$=1.20$\mu_0$ ↑ | *Nontrivial* (4, 4,9) | - |
| | 2NN Stretching $\mu$=0.80$\mu_0$ ↓ | Trivial | - |
| | Bending and bending-stretching coupling ($\beta, \gamma, \xi$)=1.2($\beta, \gamma, \xi$)$_0$ ↑ | *Nontrivial* (2, 4) | - |
| | Bending and bending-stretching coupling ($\beta, \gamma, \xi$)=0.8($\beta, \gamma, \xi$)$_0$ ↓ | *Nontrivial* (4, 4,7) | - |
| SW | Original | *Nontrivial* (2, 1) | *Nontrivial* (4, 1,4) |
| | Two-body $A$=1.2$A_0$ ↑ | *Nontrivial* (2, 1) | *Nontrivial* (2, 4) |
| | Two-body $A$=0.8$A_0$ ↓ | *Nontrivial* (4, 1,4) | *Nontrivial* (6, 1,4,9) |
| | Three-body $\lambda$=1.2$\lambda_0$ ↑ | *Nontrivial* (4, 1,4) | *Nontrivial* (8, 1,4,9,11) |
| | Three-body $\lambda$=0.8$\lambda_0$ ↓ | *Nontrivial* (2, 1) | *Nontrivial* (2, 4) |
| Tersoff | Original | Trivial | Trivial |
| | Repulsive $A$=1.01$A_0$↑ | *Nontrivial* (2, 9) | Trivial |
| | Repulsive $A$=0.99$A_0$ ↓ | *Nontrivial* (2, 10) | Trivial |
| | Attractive $B$=1.01$B_0$ ↑ | *Nontrivial* (2, 10) | Trivial |
| | Attractive $B$=0.99$B_0$ ↓ | Trivial | Trivial |

Table 4. Weyl phonon properties of GaN and AlN with varying coefficients in nonlocality. The numbers in parentheses denote the number of Weyl phonons in the IBZ and their band index.

| Models \ Materials | Variations of coefficients | GaN | AlN |
|---|---|---|---|
| KVFF | Original | *Nontrivial* (4, 1,4) | *Nontrivial* (4, 1,4) |
| | 3NN (small, $\delta=\delta_0$) | *Nontrivial* (4, 1,4) | *Nontrivial* (4, 1,4) |
| | 3NN (medium, $\delta=2\delta_0$) | *Nontrivial* (4, 1,4) | *Nontrivial* (2, 4) |
| | 3NN (large, $\delta=5\delta_0$) | *Nontrivial* (2, 4) | *Nontrivial* (4, 4,10) |
| | Real 3NN (small, $\delta=\delta_0$, $\delta'=\delta'_0$) | - | *Nontrivial* (2, 4) |
| | Real 3NN (large, $\delta=\delta_0$, $\delta'=2\delta'_0$) | - | *Nontrivial* (6, 4,9,10) |
| OVFF | Original | *Nontrivial* (2, 4) | - |
| | 3NN (small, $\delta=\delta_0$) | *Nontrivial* (2, 4) | - |
| | 3NN (medium, $\delta=10\delta_0$) | *Nontrivial* (2, 4) | - |
| | 3NN (large, $\delta=20\delta_0$) | *Nontrivial* (2, 4) | - |
| | Without 2NN | Trivial | - |
| | Without COP | Trivial | - |
| SW | Original | *Nontrivial* (2, 1) | *Nontrivial* (4, 1,4) |
| | 3NN (std., $\varepsilon=\varepsilon_0$) | Trivial | *Nontrivial* (4, 1,4) |
| | 3NN (small, $\varepsilon=0.1\varepsilon_0$) | Trivial | *Nontrivial* (4, 1,4) |
| | 3NN (medium, $\varepsilon=2\varepsilon_0$) | *Nontrivial* (4, 1,4) | *Nontrivial* (2, 4) |
| | 3NN (large, $\varepsilon=10\varepsilon_0$) | *Nontrivial* (2, 4) | *Nontrivial* (6,4,9,10) |
| Tersoff | Original | Trivial | Trivial |
| | 3NN (std., $\varepsilon=\varepsilon_0$) | Trivial | *Nontrivial* (2, 4) |
| | 3NN (small, $\varepsilon=0.1\varepsilon_0$) | *Nontrivial* (4, 9,10) | Trivial |
| | 3NN (medium, $\varepsilon=2\varepsilon_0$) | Trivial | *Nontrivial* (2, 4) |
| | 3NN (large, $\varepsilon=10\varepsilon_0$) | *Nontrivial* (2, 9) | Trivial |

## Figure captions

Figure 1. Schematics of wurtzite GaN and AlN, and nearest neighbor atoms.

Figure 2. Phonon dispersions of wurtzite GaN with (a) IFC models and (b) potential functions.

Figure 3. Phonon dispersions of wurtzite AlN with (a) IFC models and (b) potential functions.

Figure 4. Nontrivial phonon crossing points by IFC models and potential functions

Figure 5. Nontrivial phonon crossing points by IFC models under conditions with varying IFC strengths

Figure 6. Nontrivial phonon crossing points by potential functions under conditions with varying IFC strengths

Figure 7. Positions of Weyl phonons in IBZs under conditions with varying coefficients of IFC strength

Figure 8. Nontrivial phonon crossing points modeled by potentials under conditions with varying IFC nonlocality

Figure 9. Nontrivial phonon crossing points modeled by potentials under conditions with varying IFC nonlocality

Figure 10. Positions of Weyl phonons in IBZs under conditions with varying coefficients of IFC nonlocality

Figure 11. (a),(b),(e),(f),((i),(j) Band inversions in Weyl phonon phase transition of AlN modeled by different IFC models with 3NN interactions and corresponding (c),(d),(g),(h),(k),(l) Berry curvature distributions.

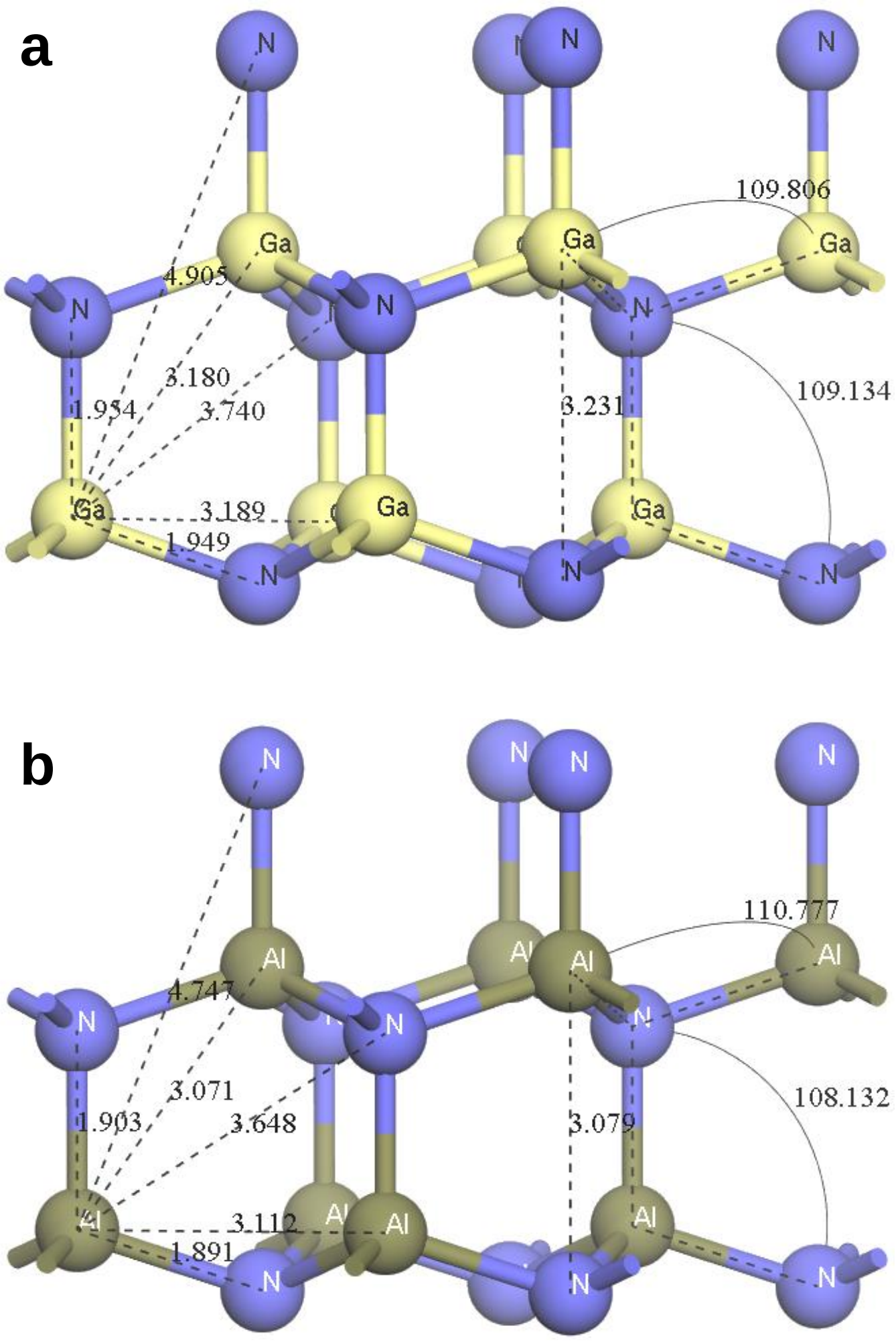

Figure 1. Schematics of wurtzite GaN and AlN, and nearest neighbor atoms.

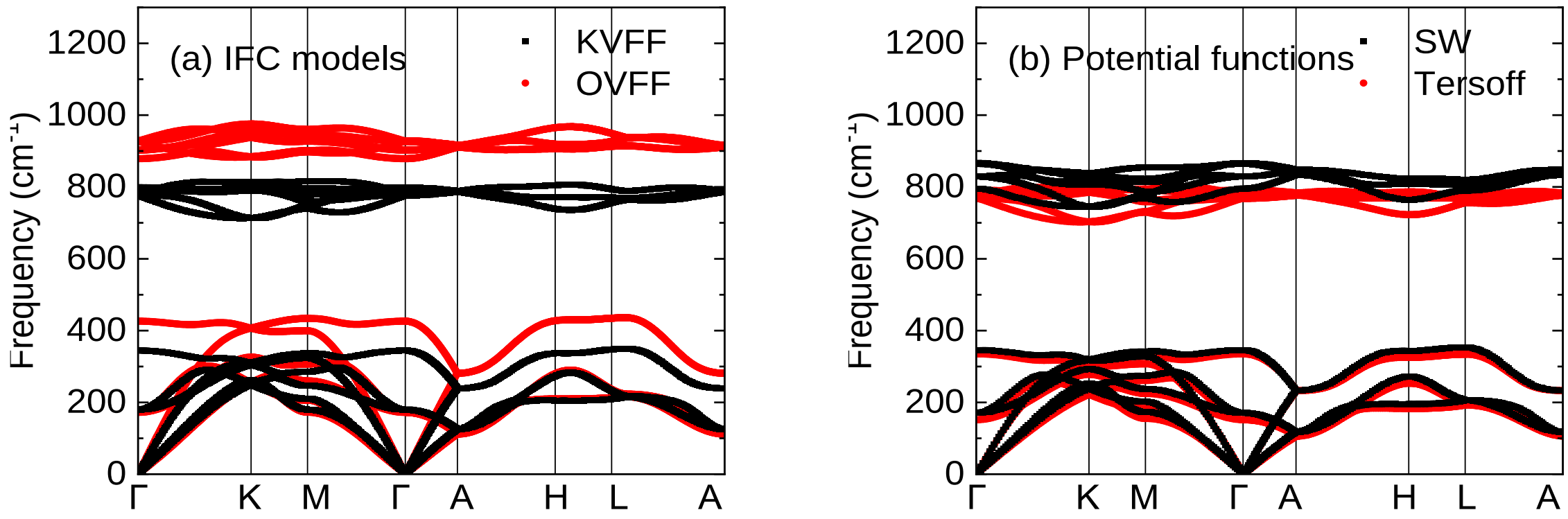

Figure 2. Phonon dispersions of wurtzite GaN with (a) IFC models and (b) potential functions.

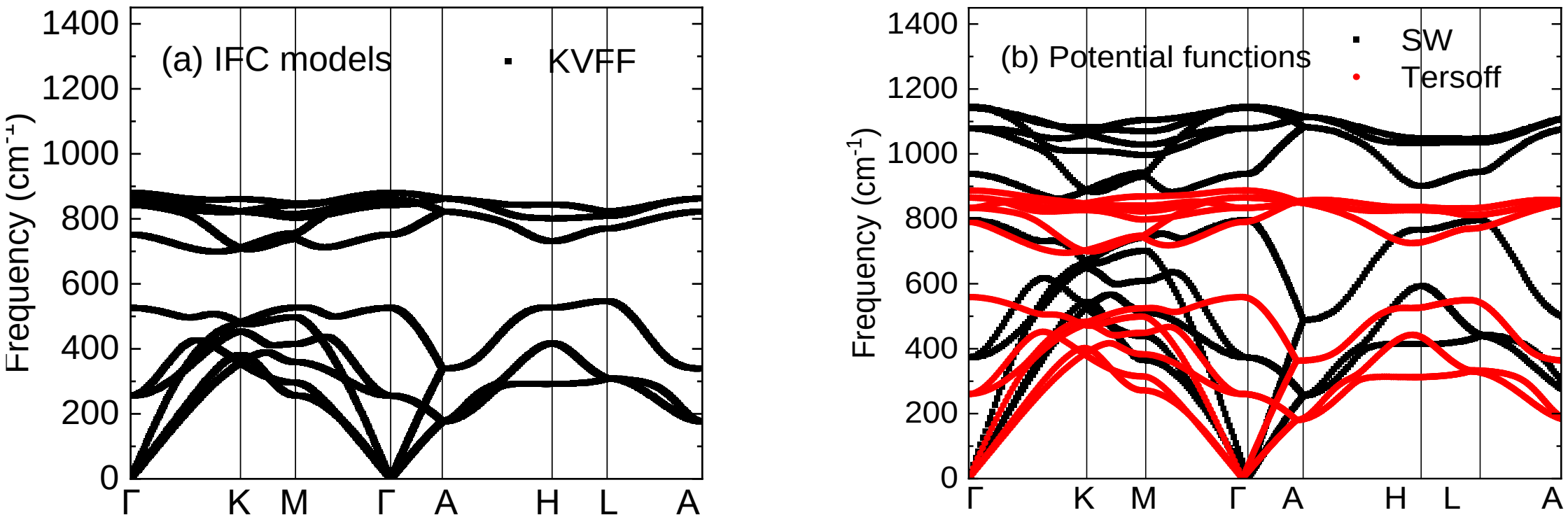

Figure 3. Phonon dispersions of wurtzite AlN with (a) IFC models and (b) potential functions.

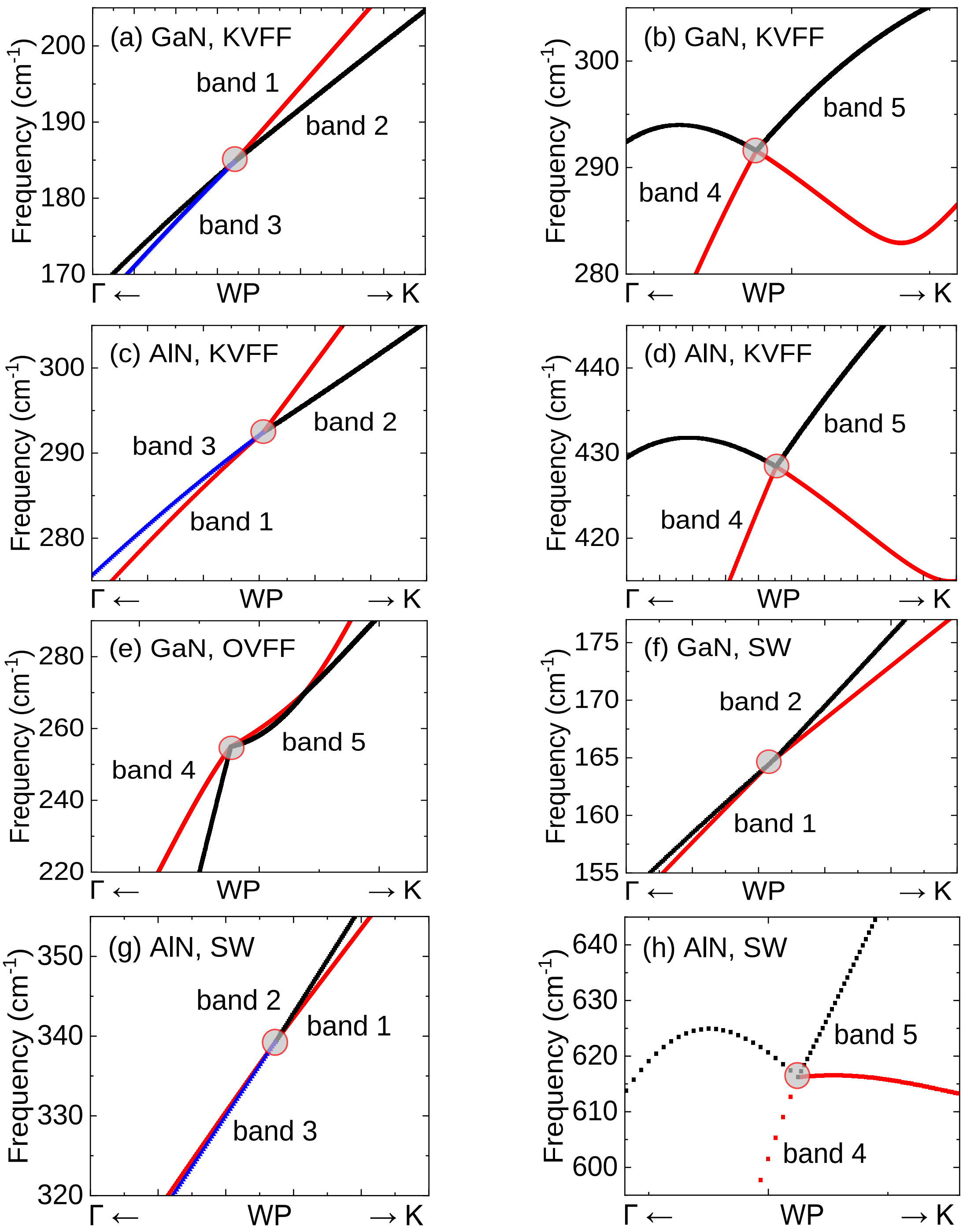

Figure 4. Nontrivial phonon crossing points by IFC models and potential functions

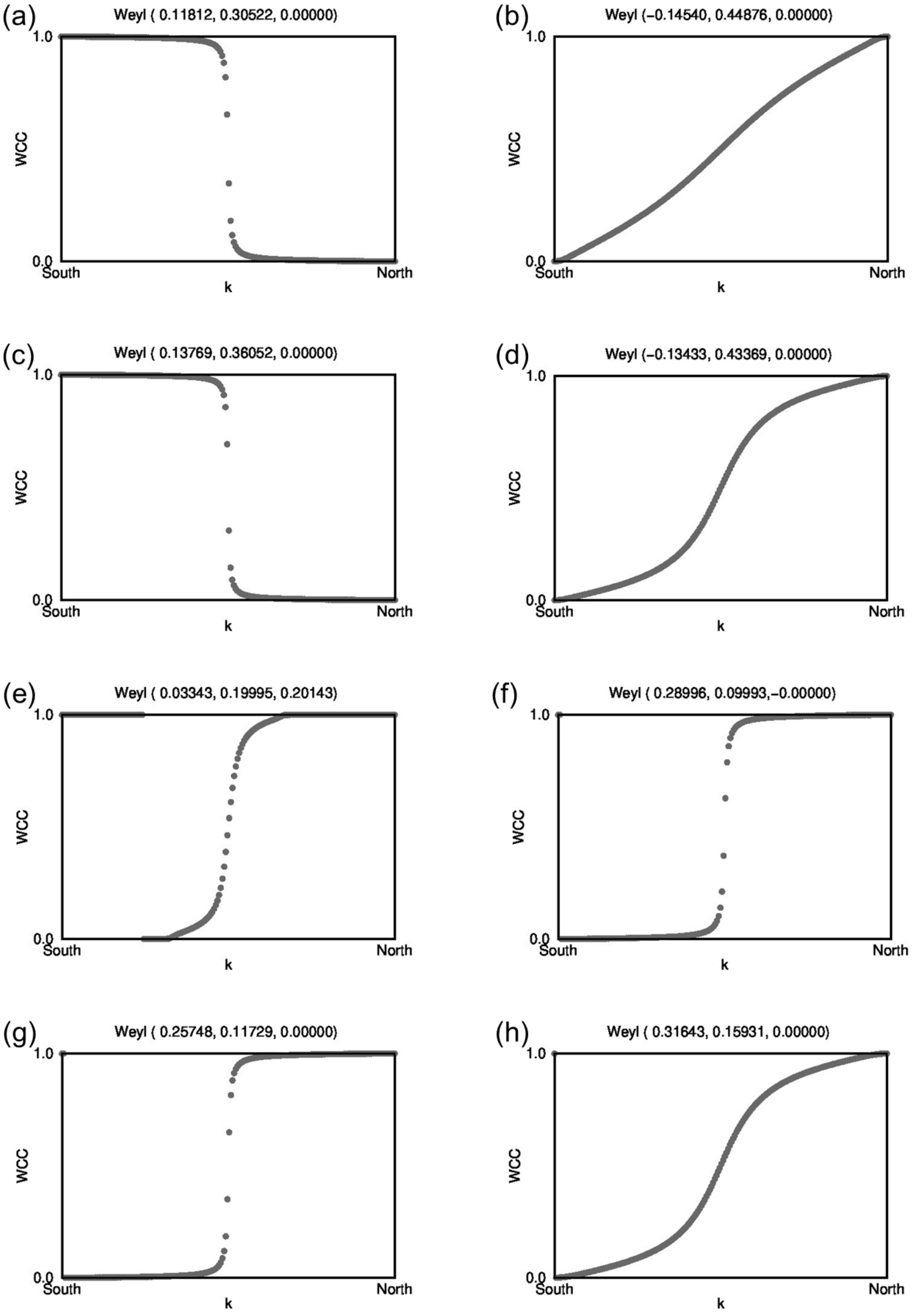

Figure 5. Corresponding topological features in original models to Figure 4: WCC evolution of nontrivial phonon crossing points by IFC models and potential functions

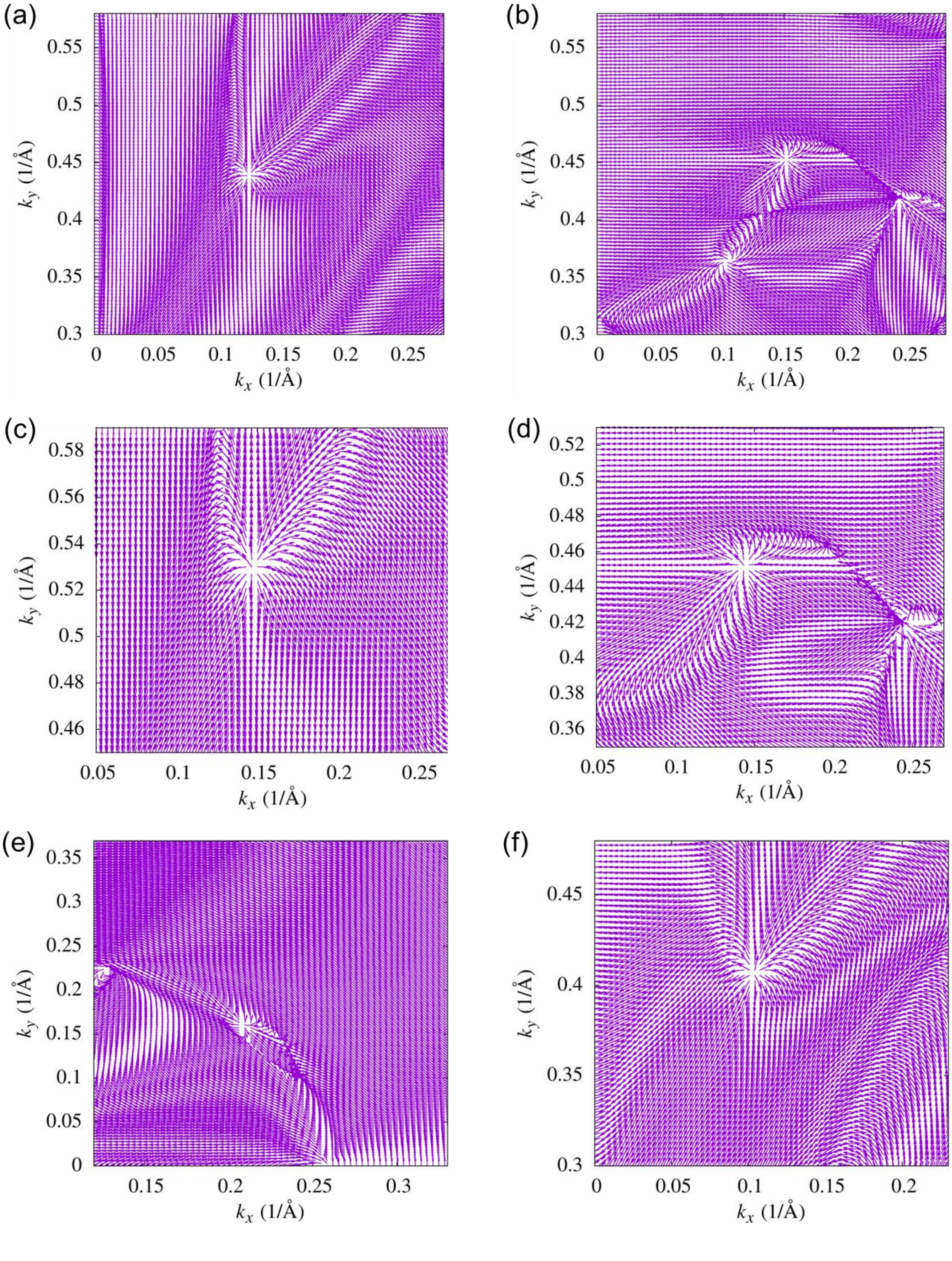

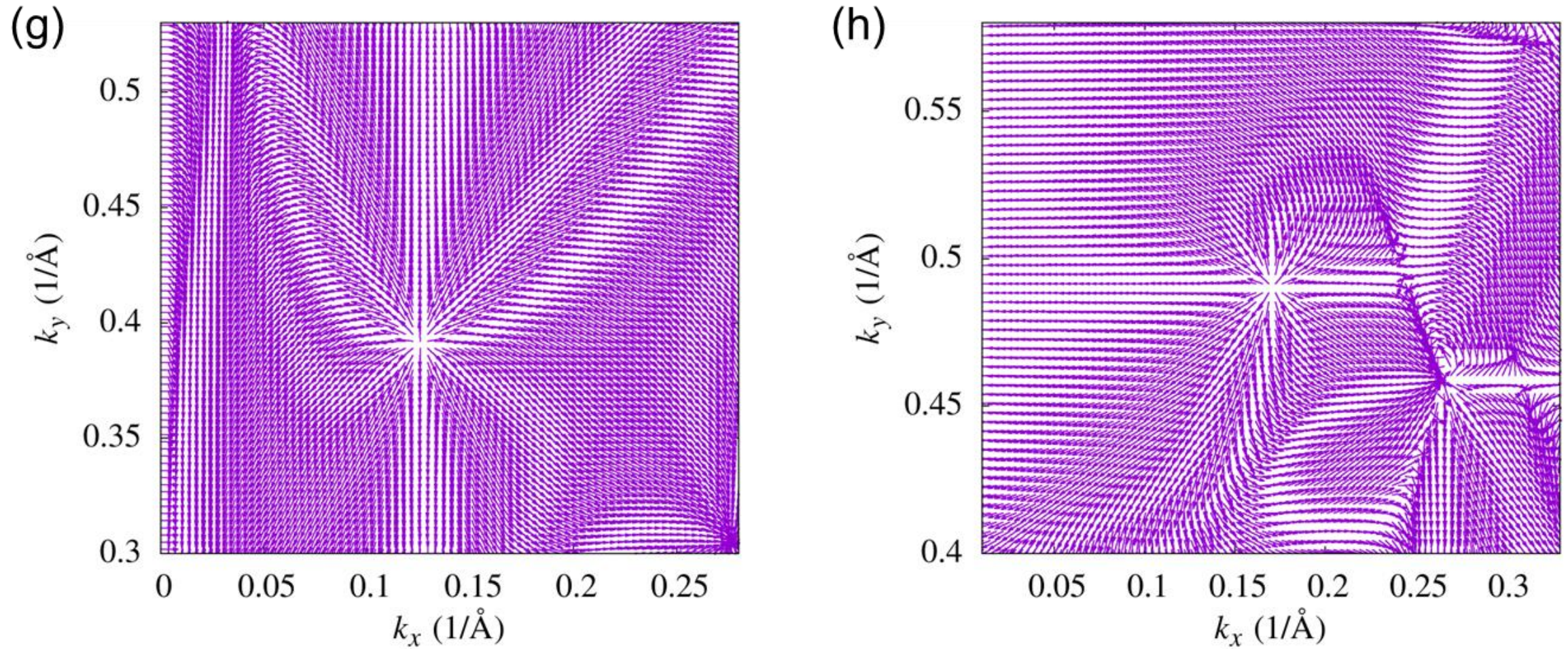

Figure 6. Corresponding topological features in original models to Figure 4: Berry curvature of nontrivial phonon crossing points by IFC models and potential functions

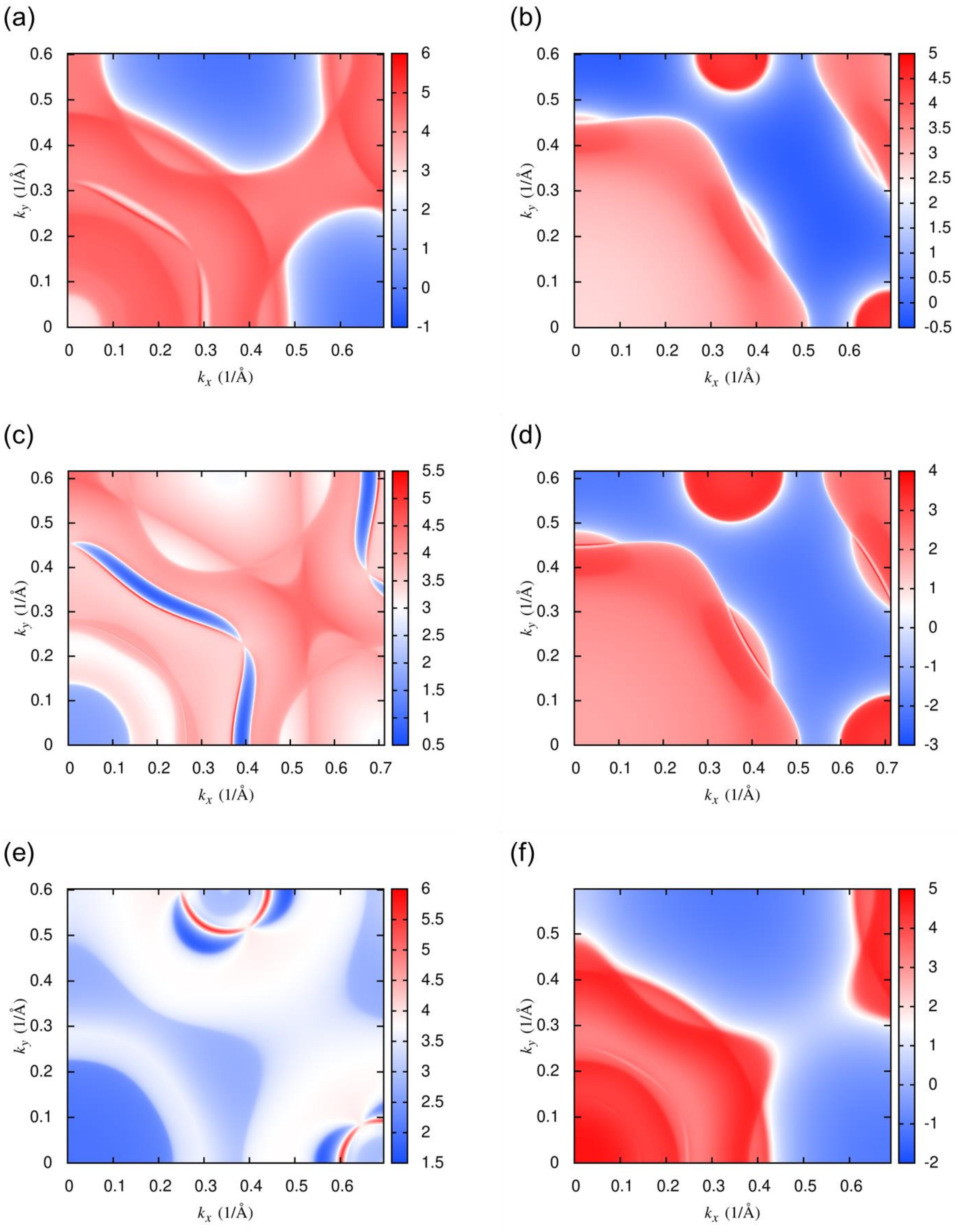

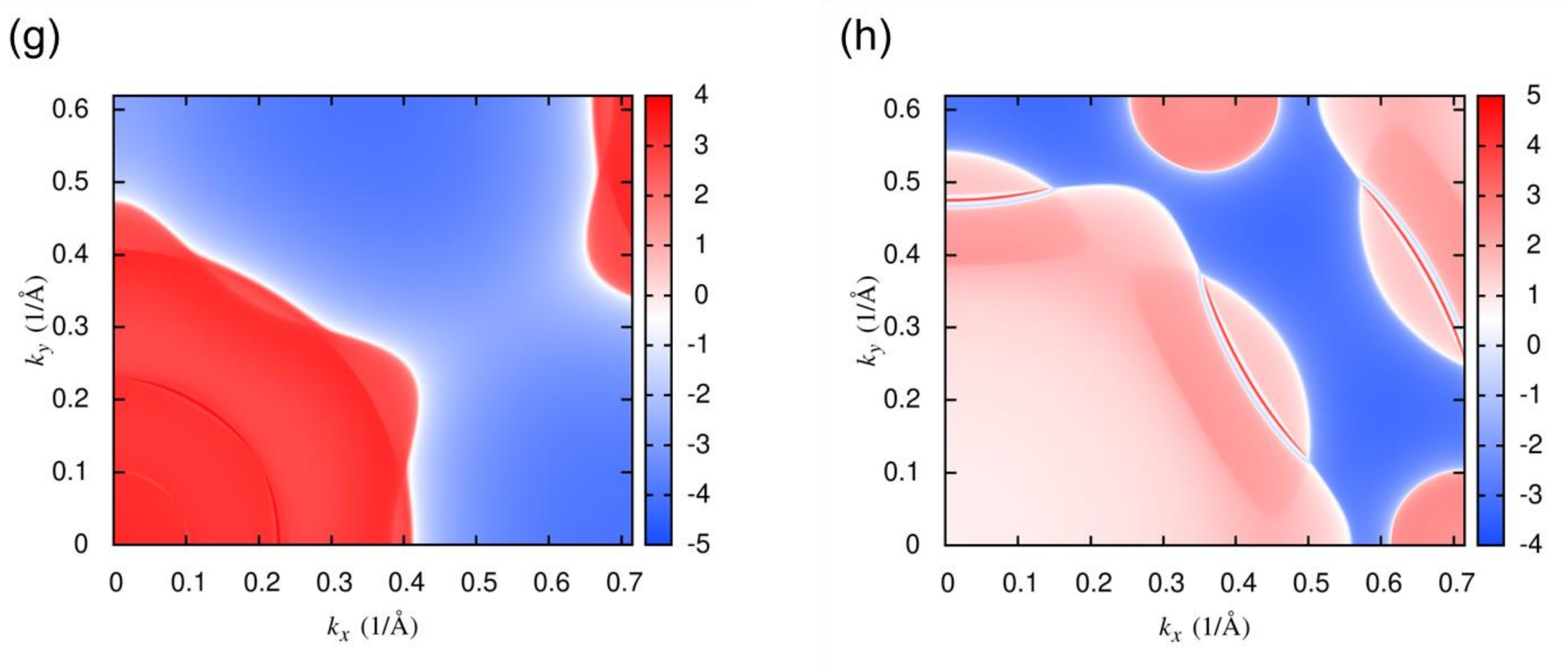

Figure 7. Corresponding topological features in original models to Figure 4: arc surface states of nontrivial phonon crossing points by IFC models and potential functions

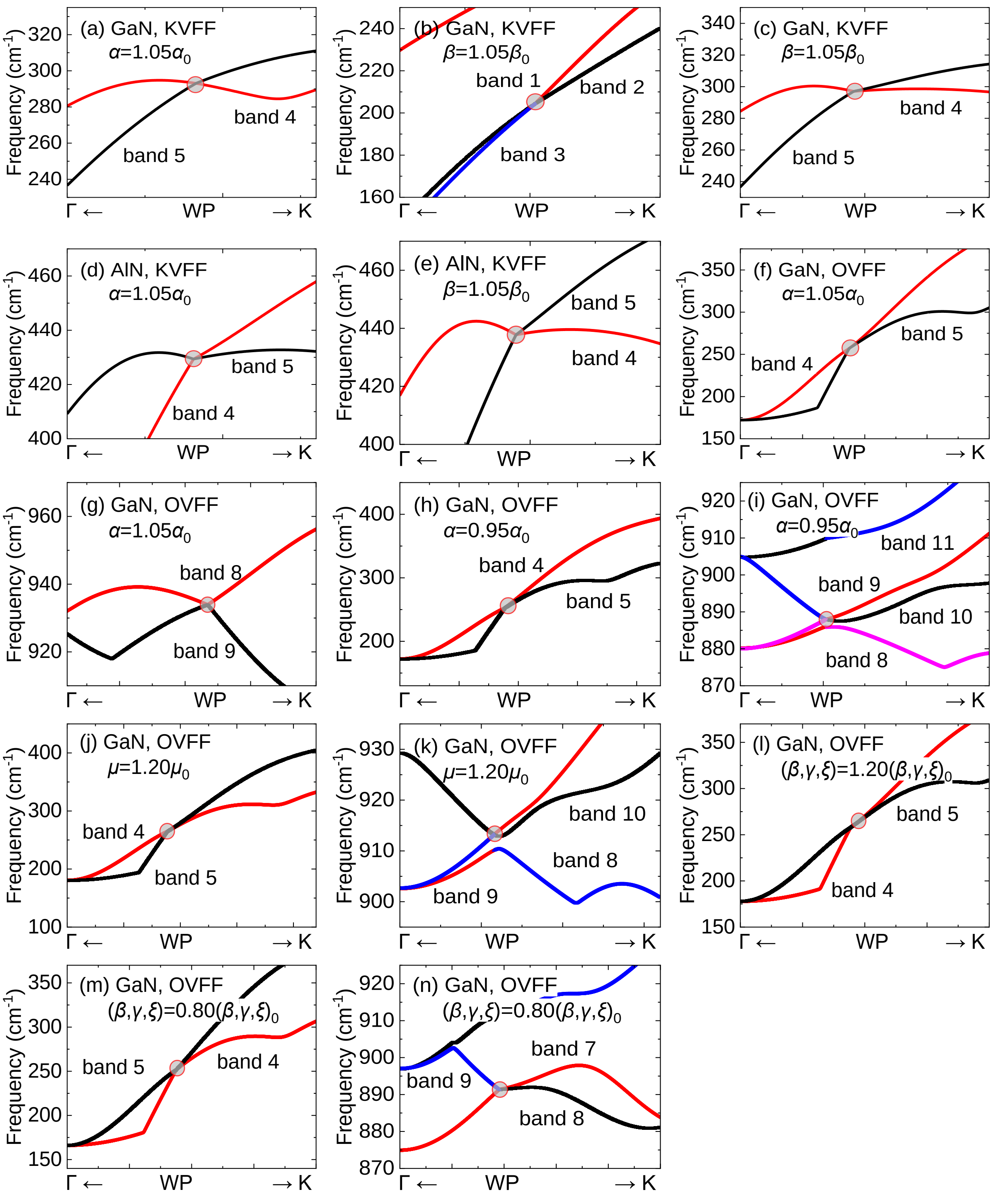

Figure 8. Nontrivial phonon crossing points by IFC models under conditions with varying IFC strengths

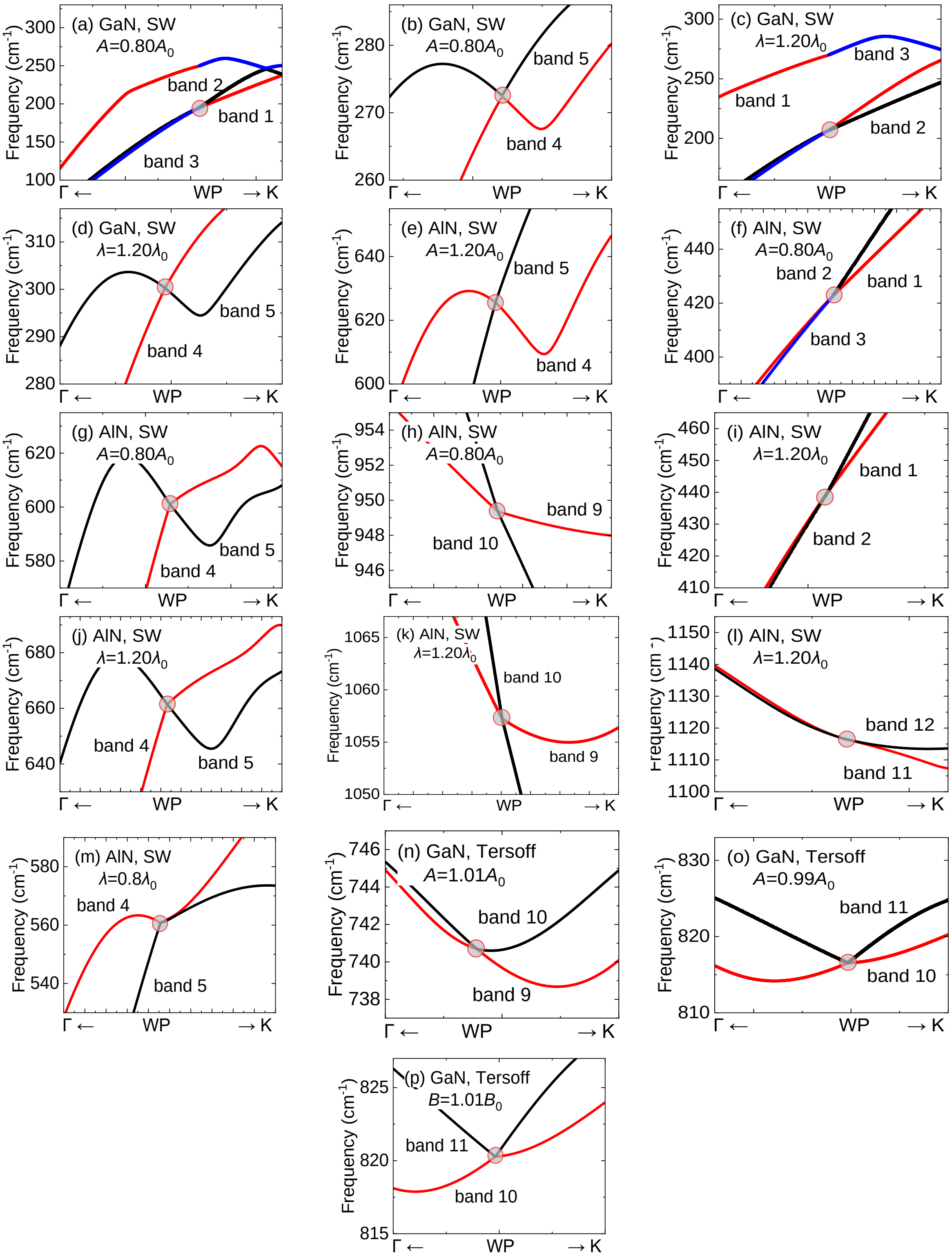

Figure 9. Nontrivial phonon crossing points by potential functions under conditions with varying IFC strengths

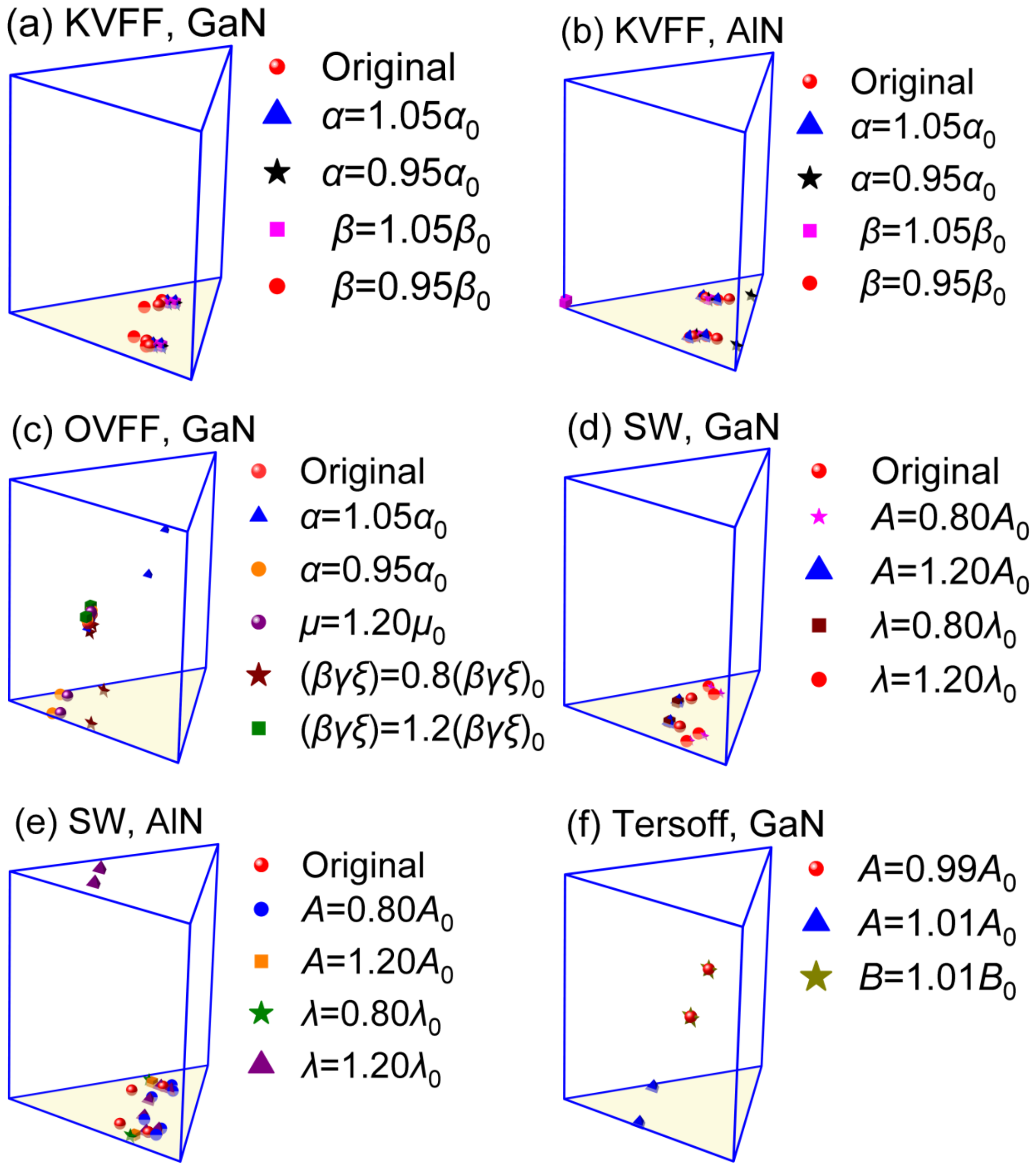

Figure 10. Positions of Weyl phonons in IBZs under conditions with varying coefficients of IFC strength

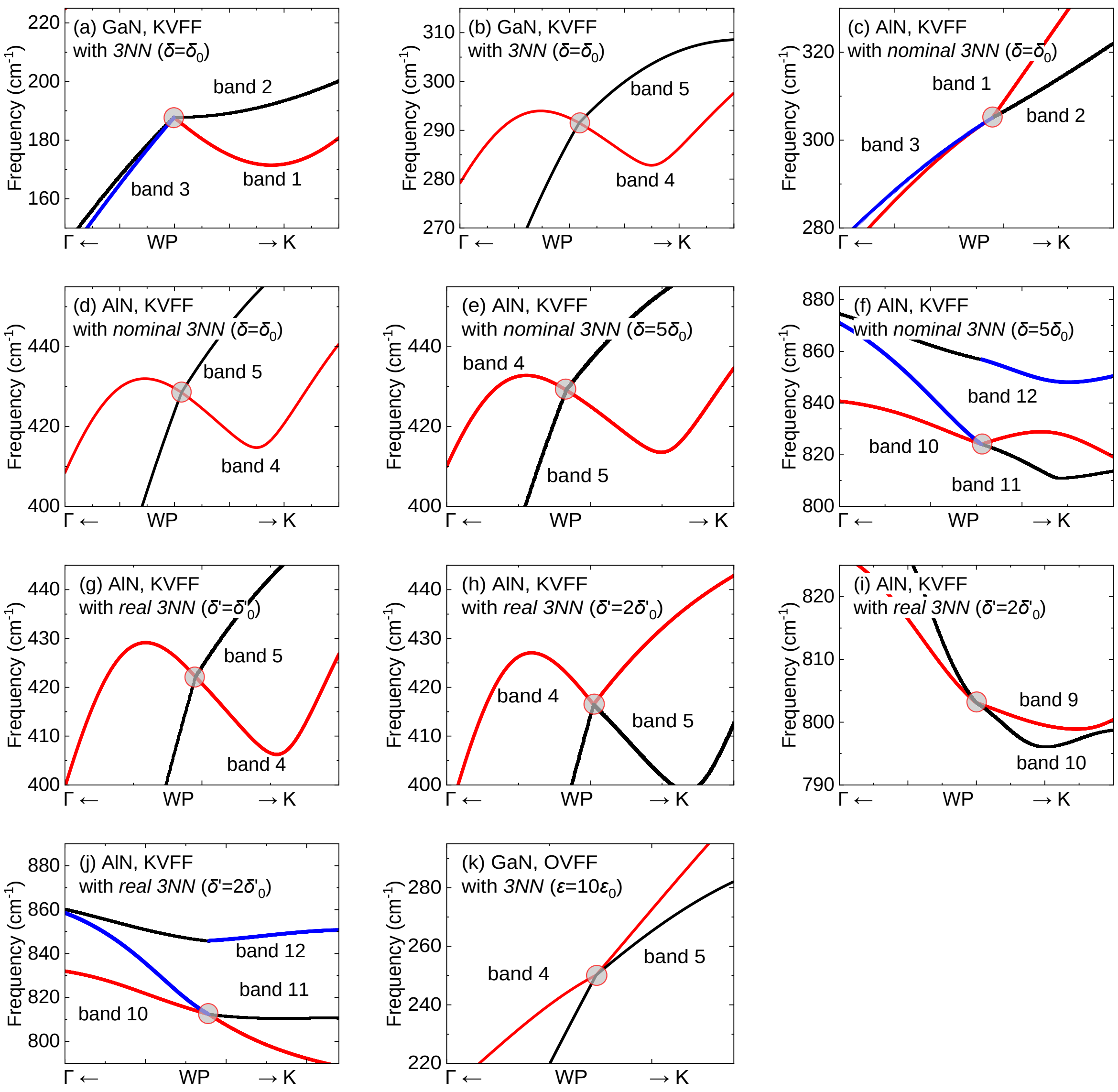

Figure 11. Nontrivial phonon crossing points modeled by potentials under conditions with varying IFC nonlocality

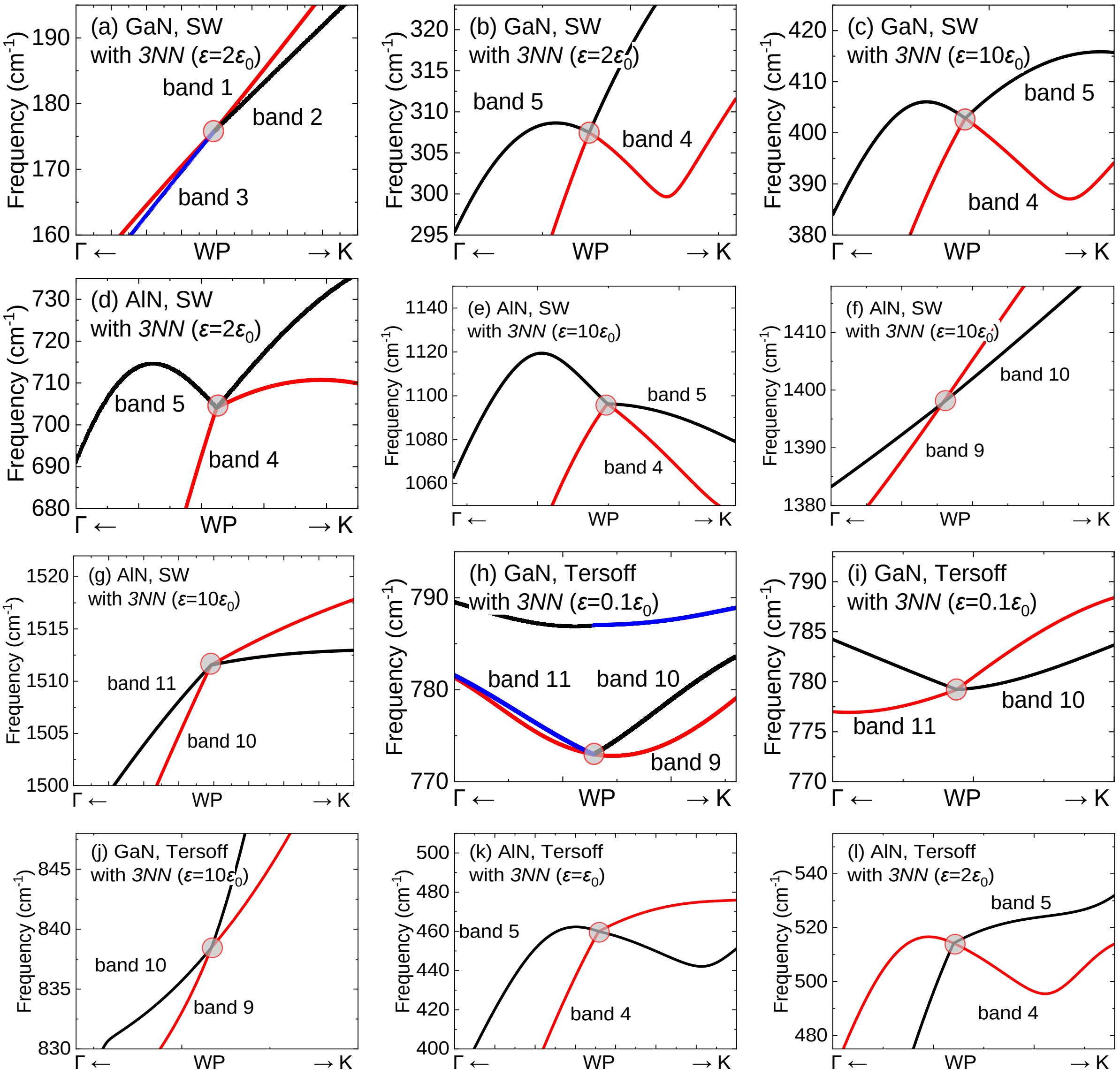

Figure 12. Nontrivial phonon crossing points modeled by potentials under conditions with varying IFC nonlocality

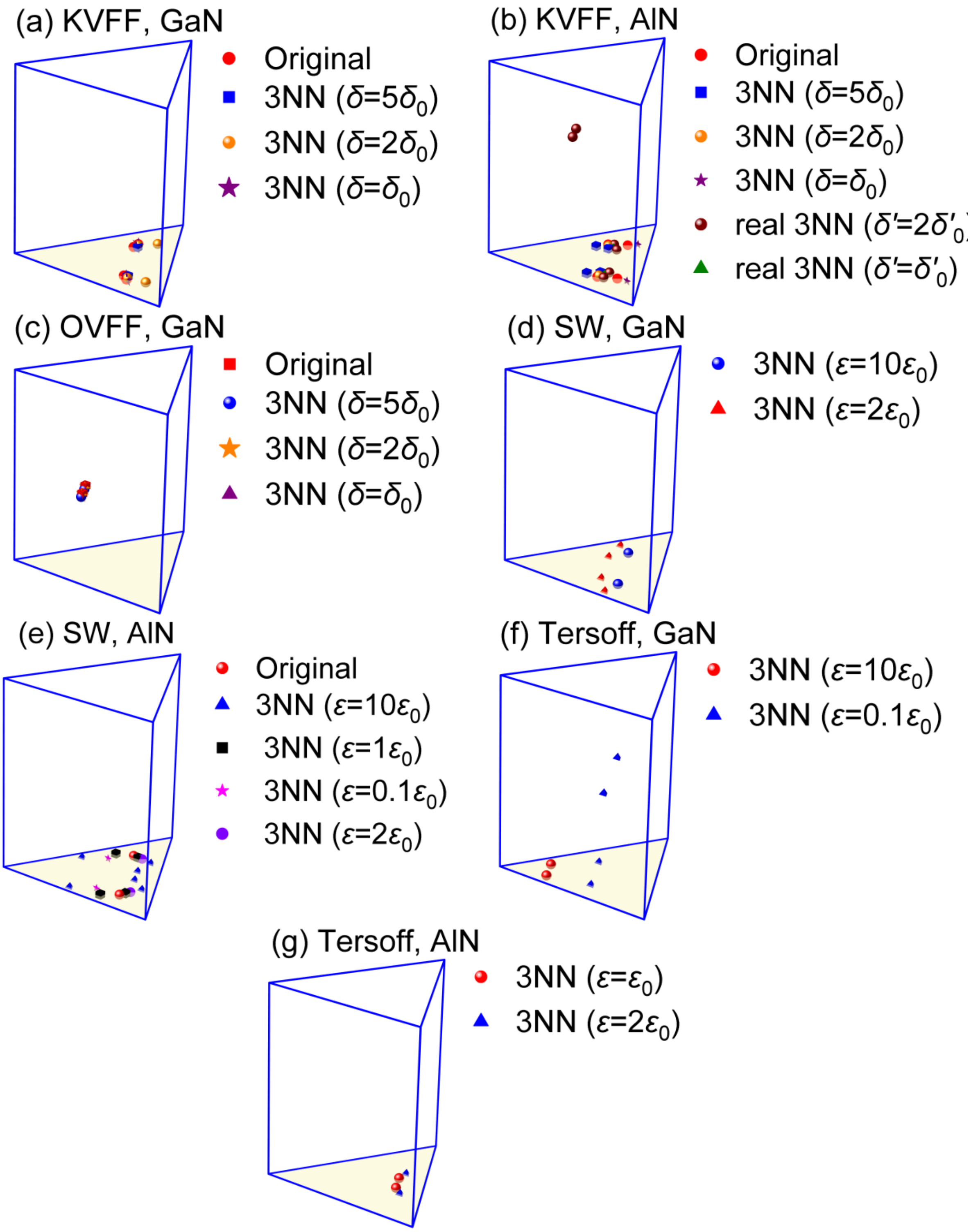

Figure 13. Positions of Weyl phonons in IBZs under conditions with varying coefficients of IFC nonlocality

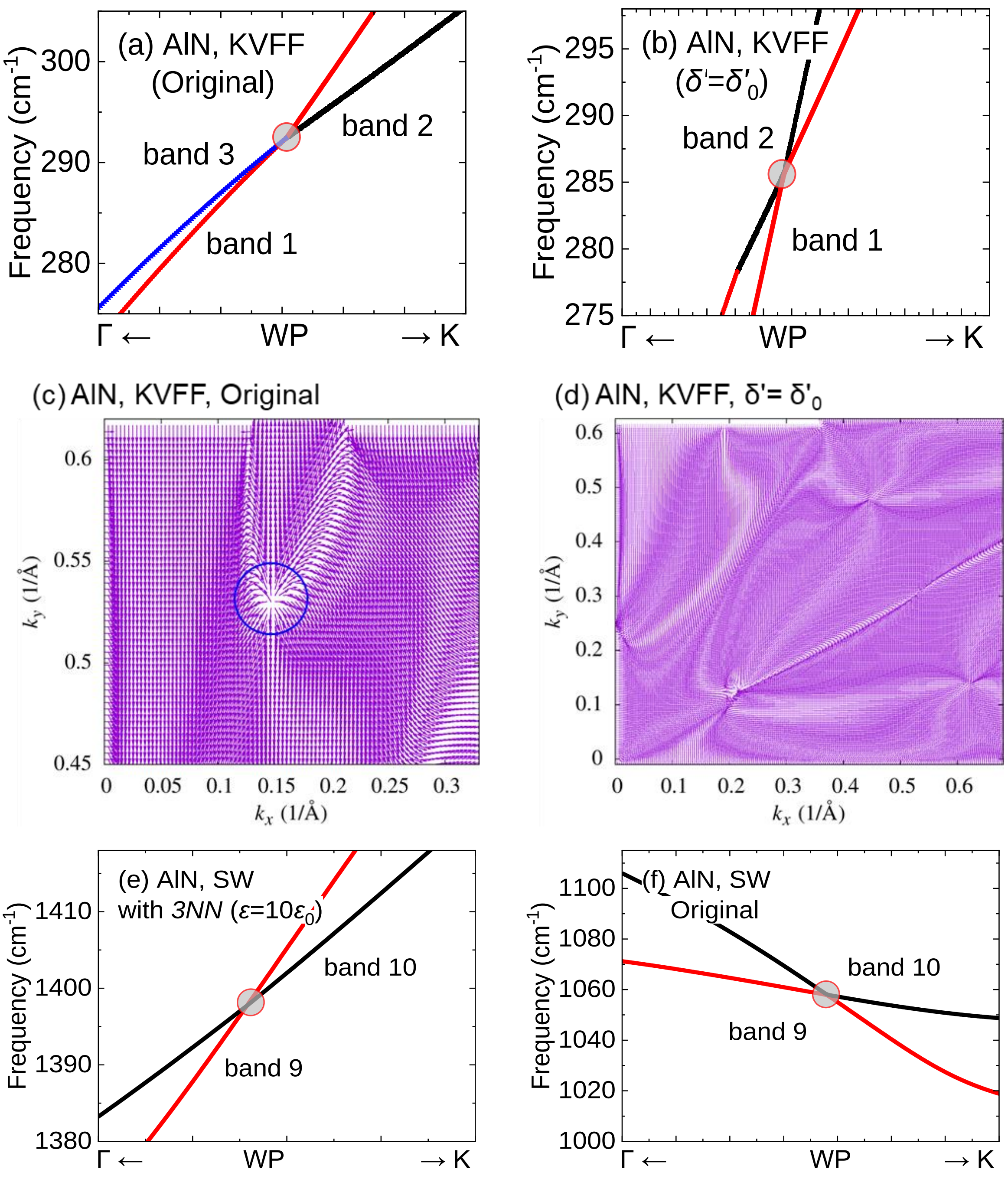

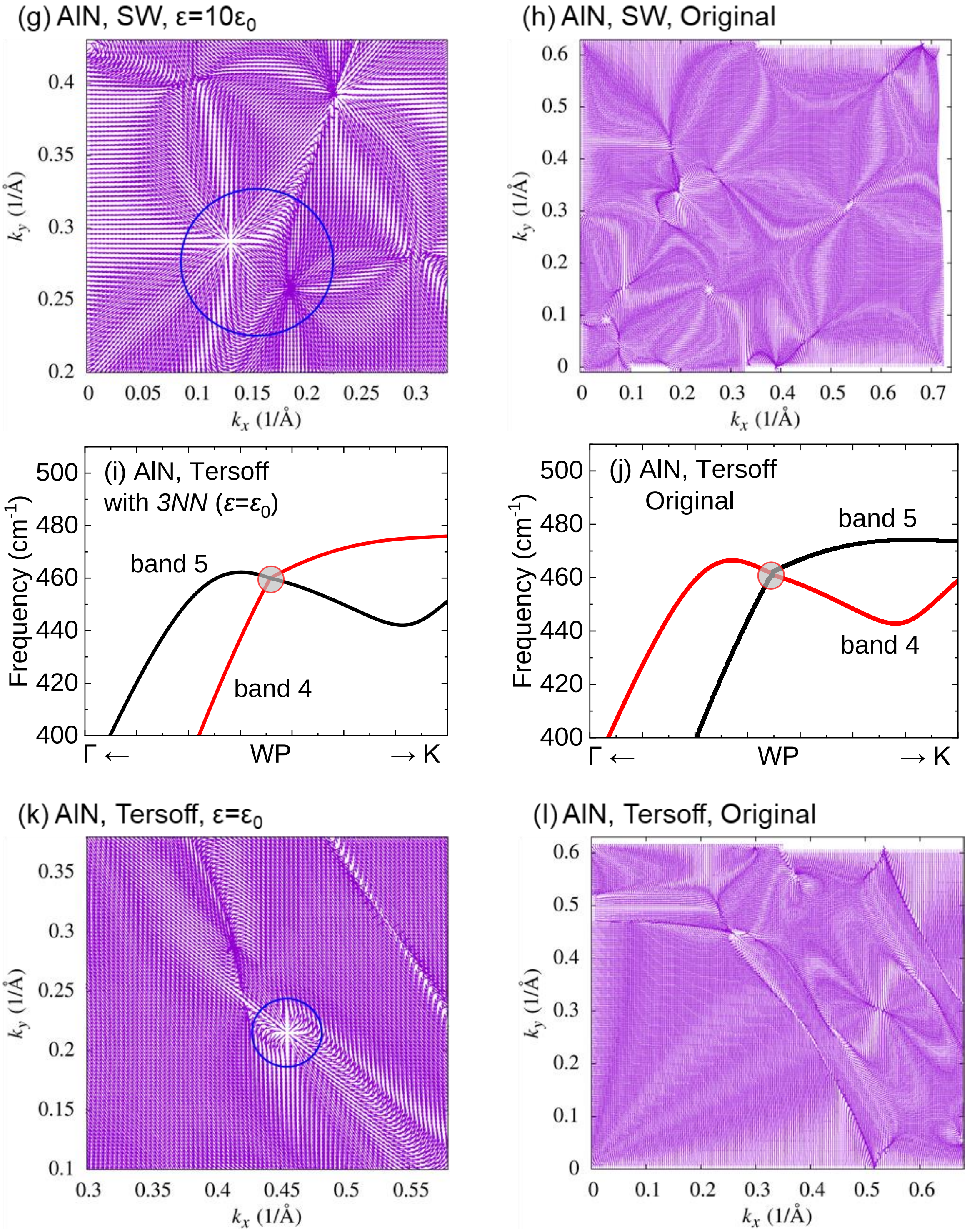

Figure 14. (a),(b),(e),(f),((i),(j) Band inversions in Weyl phonon phase transition of AlN modeled by different IFC models with 3NN interactions and corresponding (c),(d),(g),(h),(k),(l) Berry curvature distributions.